\documentclass[12pt,preprint]{aastex}
\usepackage{natbib}
\usepackage{amssymb}
\usepackage{epsfig}
\usepackage{lscape}
\usepackage{longtable}

\newcommand{\etal}{et~al.~}
\newcommand\tna{\,\tablenotemark{a}}
\newcommand\tnb{\,\tablenotemark{b}}

\begin{document}


\title{On the Origin of the Extended H$\alpha$ Filaments in \\Cooling Flow Clusters}

\author{Michael McDonald\altaffilmark{1,3}, Sylvain Veilleux\altaffilmark{1,4}, David S.N. Rupke\altaffilmark{2} and Richard Mushotzky\altaffilmark{1}}

\altaffiltext{1}{Astronomy Department, University of Maryland, College
  Park, MD 20742} 
\altaffiltext{2}{Institute for Astronomy, University of Hawaii, 2680 Woodlawn Dr., Honolulu, HI 96822, USA}
\altaffiltext{3}{Email: mcdonald@astro.umd.edu}
\altaffiltext{4}{Email: veilleux@astro.umd.edu}


\begin{abstract}

We present a high spatial resolution H$\alpha$ survey of 23 cooling
flow clusters using the Maryland Magellan Tunable Filter (MMTF),
covering 1--2 orders of magnitude in cooling rate, dM/dt, temperature
and entropy. We find 8/23 (35\%) of our clusters have complex,
filamentary morphologies at H$\alpha$, while an additional 7/23 (30\%)
have marginally extended or nuclear H$\alpha$ emission, in general
agreement with previous studies of line emission in cooling flow
cluster BCGs. A weak correlation between the integrated near-UV
luminosity and the H$\alpha$ luminosity is also found for our complete
sample, with a large amount of scatter about the expected relation for
photoionization by young stars. We detect H$\alpha$ emission out to
the X-ray cooling radius, but no further, in several clusters and find
a strong correlation between the H$\alpha$ luminosity contained in
filaments and the X-ray cooling flow rate of the cluster, suggesting
that the warm ionized gas is linked to the cooling flow. Furthermore,
we detect a strong enhancement in the cooling properties of the ICM
coincident with the H$\alpha$ emission, compared to the surrounding
ICM at the same radius. While the filaments in a few clusters may be
entrained by buoyant radio bubbles, in general, the radially-infalling
cooling flow model provides a better explanation for the observed
trends. The correlation of the H$\alpha$ and X-ray properties suggests
that conduction may be important in keeping the filaments ionized. The
thinness of the filaments suggests that magnetic fields are an
important part of channeling the gas and shielding it from the
surrounding hot ICM.

\end{abstract}



\section{Introduction}
The high densities and low temperatures of the intracluster medium
(hereafter ICM) in the cores of some galaxy clusters imply massive
amounts of gas cooling radiatively out of the ICM. Early studies (see
review by Fabian 1994) suggested that cooling flows on the order of
100--1000 M$_{\odot}$~yr$^{-1}$ should be depositing massive amounts
of cold gas onto the brightest cluster galaxy (hereafter BCG) in the
cluster core. The lack of evidence for such large amounts of cold
molecular gas or young stars is often referred to as the ``cooling
flow problem''.  More recent studies suggest that the cooling region
is much smaller than previously thought (Peres \etal 1998), which
implies overall lower estimates on the mass deposition
rate. Additionally, it is now generally accepted that some form of
heating balances the radiative cooling, leading to cooling flow rates
on the order of 1--10 M$_{\odot}$~yr$^{-1}$. Mechanical and radiative
feedback from AGN and starbursts (e.g. Peterson \& Fabian 2006,
Veilleux, Cecil \& Bland-Hawthorn 2005), gas sloshing in the cluster
core (e.g. Zuhone \etal 2009), merger shocks (e.g. Randall \etal 2002)
and conduction from the surrounding ICM (e.g. Voigt \& Fabian 2004)
have all been suggested as possible contributors to ICM
heating. However, the relative importance of each process remains
unclear.

The presence of optical line-emitting nebulae, found primarily in
cooling flow clusters, is a clue that cooler ($\sim$~10$^4$~K) gas
coexists with the ICM and that cooling flows may not be entirely ruled
out (Hu \etal 1985, Heckman \etal 1989, Crawford \etal 1999; Jaffe
\etal 2005, Hatch \etal 2007). However, while several intriguing ideas
have been put forward, the origin of this gas and the mechanism for
heating it are, as yet, unknown. While the presence of cool gas could
be attributed to the purported cooling flow, one would then naively
expect a relatively symmetric geometry and a strong connection of
optical emission lines with X-ray properties. However, this is not
generally the case. Crawford \etal (2005) investigate a specific
scenario, the extended H$\alpha$ filament in Abell~1795, and suggest
as an explanation a cooling wake, produced by a symmetric cooling flow
falling onto the brightest cluster galaxy (hereafter BCG) which has a
high relative velocity at the center of the cluster. Oegerle \& Hill
(2001) find that $\sim$ 15\% of BCGs are moving with a high peculiar
velocity, so an asymmetric distribution of warm gas in this fraction
of clusters would be relatively straightforward to explain in this
context. Potentially contributing heating/ionization sources in the
filaments include (i) the central AGN, (ii) hot young stellar
population outside the cD galaxy nucleus, (iii) X-rays from the ICM
itself, (iv) heat conduction from the ICM to the colder filaments, (v)
shocks and turbulent mixing layers, and (vi) collisional heating by
cosmic rays (Crawford \etal 2005; Ferland \etal 2009).

In star forming regions, the presence of ionized hydrogen is a prime
indication of the rate and location of star formation. While several
cooling flow clusters have been observed to have non-zero star
formation rates at their center, based on UV (Hicks \& Mushotzky 2005)
and infrared (O'Dea \etal 2008) observations, it is not immediately
clear whether newly formed stars are fully responsible for the flux
and morphology of the observed H$\alpha$ emission. Indeed, there has
been very little work done on the spatial correlation between the star
forming and warm components in cooling flow cluster centers. What
\emph{has} become clear from recent studies is that, in general, the
measured star formation rate in the BCG is consistently less than the
spectroscopically determined X-ray cooling rate for the cluster by a
factor of 3-10 (i.e. Rafferty \etal 2006; O'Dea \etal 2008), implying
that not all of the cooling gas is turning into stars with a Salpeter
(1955) IMF.

A major problem standing in the way of an explanation of these
filaments is that there are still only a small number of known cases
with clearly extended emission beyond the nucleus of the BCG. This is
due to the paucity of narrow-band filters which can be tuned to
cosmological redshifts. Attempts have been made to quantify the
fraction of BCGs which exhibit H$\alpha$ emission using the Sloan
Digital Sky Survey (e.g. Edwards \etal\ 2007). However, spectroscopic
studies usually only tell us about the very center of the BCG, which
can be bright in H$\alpha$ if an AGN is present. What is
needed is a deep, wide-field, narrow-band survey of clusters of
various richness in order to assess the ubiquity, extent, and topology
of the warm gas. Some progress has been made towards this end by Jaffe
\etal (2005), using the Taurus Tunable Filter (TTF; Bland-Hawthorn \&
Jones 1998). The use of a tunable filter allows the observer to tune
to a large range in wavelengths with a narrow bandwidth
($\sim$~10\AA). The ability to tune the wavelength is particularly
useful when observing cooling flow clusters, of which there are very
few in the local Universe.

In the pilot study leading to this survey paper, McDonald \& Veilleux
(2009; hereafter MV09) observed Abell~1795 in H$\alpha$ with the
Maryland Magellan Tunable Filter (hereafter MMTF; Veilleux \etal 2010)
in H$\alpha$. The well-known southeast filament which extends
$\sim$~50 kpc from the BCG was found to be, in fact, two separate thin
strands. These thin filaments were resolved into chains of point
sources in the far-UV with the \emph{Hubble Space Telescope}
(\emph{HST}) Advanced Camera for Surveys Solar Blind Channel. We
suggested that these strands of H$\alpha$ were, in fact, a cooling
wake produced by the cooling flow falling onto a cD galaxy which is
plowing through the center of the cluster. As the gas cools, it
collapses into thinner filaments, eventually leading to star formation
which will ionize the gas. The thinness of the filaments suggests that
magnetic fields are likely important to prevent turbulence from
erasing them, but this is just speculation without any direct evidence
of a strong magnetic field in the filaments. Although it is likely
more complicated than this (we observe higher-than- expected [N
II]/H$\alpha$ ratios, and a slight spatial offset between the UV and
H$\alpha$), this scenario can explain the general structure of the
H$\alpha$ complex.

In order to have a more complete picture of the extended H$\alpha$
emission in galaxy clusters, we have undertaken a survey of 23
clusters with a wide variety of properties using the MMTF. In the
following section we describe our sample selection, as well as
the acquisition and analysis of the data. In \S3, we present the
results of the analysis and in \S4 discuss the various
implications of these results. Finally, in \S5 we present a summary of the results. 
Throughout this paper, we assume the
following cosmological values: H$_0$ = 73 km s$^{-1}$ Mpc$^{-1}$,
$\Omega_{matter}$ = 0.27, $\Omega_{vacuum}$ = 0.73.


\section{Data Collection and Analysis}
In order to have a more complete picture of cool core clusters, we
have undertaken a multi-wavelength survey of 23 cooling flow
clusters. From an initial sample size of 232 clusters, which cover 2
orders of magnitude in dM/dt and with X-ray temperatures ranging from
1-12 keV, (full sample from White \etal\ 1997) we enforced the
following cuts: $\delta < +35^{\circ}$ and $0.025 \leq z \leq 0.092$.
These cuts ensured that (1) the cluster was visible with Magellan, (2)
the appropriate blocking filters were available to observe redshifted
H$\alpha$, and (3) the cluster angular size did not exceed that of
MMTF ($\sim$30\arcmin). These two cuts removed a total of 44 and 62
clusters from the full sample, respectively, leaving a total of 126
observable clusters. From this list, we selected 23 clusters to cover
the full range of properties, from very rich clusters with high
cooling rates (e.g., Abell 1795, Abell 2029) to low-density clusters
with little or no cooling flow (e.g., Ophiuchus, Abell 3376). A
summary of these properties for our sample clusters is given in Table
\ref{sample}. As is shown in Table \ref{sample}, the range in
classical cooling rates for this sample is from 6.3--431
M$_{\odot}$~yr$^{-1}$, which means that, while we cover a large range
in properties, we are examining only cooling flow clusters. This
selection criteria should produce a higher fraction of line-emitting
BCGs compared to an un-biased cluster sample (see Edwards \etal
2007).

For these 23 clusters, we have data at H$\alpha$ (MMTF; 23/23), red
continuum (MMTF; 23/23), near-IR (2MASS; 23/23), UV (\emph{GALEX},
\emph{XMM-OM}; 21/23) and X-ray (\emph{Chandra}; 19/23). For those
clusters hosting a radio galaxy, we have also extracted VLA 1.4 GHz
fluxes from the NVSS (Condon \etal 1998) survey (17/23). The
availability of these data are summarized in Table \ref{sample}.

In the following subsections, we describe in detail the acquisition
and reduction of these data.

\begin{table}[htbp]
\caption{Sample of 23 cooling flow clusters with MMTF H$\alpha$
imaging}
\begin{center}
\begin{tabular}{c c c c c}
\\
\hline\hline
Name & z & E(B-V) & T$_X$ & \.{M}$_{class}$  \\
(1) & (2) & (3) & (4) & (5)\\
\hline
Abell 0085     & 0.0557 & 0.038 & 6.5 & 108\\
Abell 0133     & 0.0569 & 0.019 & 3.5 & 110\\
Abell 0478     & 0.0881 & 0.517 & 6.8 & 736\\
Abell 0496     & 0.0329 & 0.132 & 4.8 & 134\\
Abell 0644     & 0.0704 & 0.122 & 6.5 & 136\\
Abell 0780     & 0.0539 & 0.042 & 4.7 & 222\\
Abell 1644     & 0.0475 & 0.069 & 5.1 & 12\\
Abell 1650     & 0.0846 & 0.017 & 5.1 & 122\\
Abell 1795     & 0.0625 & 0.013 & 5.3 & 321\\
Abell 1837     & 0.0691 & 0.058 & 2.6 & 12\\
Abell 2029     & 0.0773 & 0.040 & 7.4 & 431\\
Abell 2052     & 0.0345 & 0.037 & 3.4 & 94\\
Abell 2142     & 0.0904 & 0.044 & 10.1 & 369\\
Abell 2151     & 0.0352 & 0.043 & 2.9 & 166\\
Abell 3158     & 0.0597 & 0.015 & 5.3 & 9.6\\
Abell 3376     & 0.0597 & 0.056 & 3.5 & 6.3\\
Abell 4059     & 0.0475 & 0.015 & - & -\\
Ophiuchus     & 0.0285 & 0.588 & 8.6 & 41\\
Sersic 159-03 & 0.0580 & 0.011 & 2.4 & 288\\
\\
Abell 2580\tna     & 0.0890 & 0.024 & 4.3 & 95\\
Abell 3389\tna     & 0.0267 & 0.076 & 2.0 & 22\\
\\
Abell 0970\tnb     & 0.0587 & 0.055 & 4.1 & 20\\
WBL 360-03\tnb  & 0.0274 & 0.028 & 1.8 & 10\\
\hline
\\
\end{tabular}
\end{center}
\em{}
(1) - Cluster name
\\(2) - NED redshift of BCG (http://nedwww.ipac.caltech.edu)
\\(3) - Reddening due to Galactic extinction from Schlegel \etal (1998)
\\(4) - Cluster X-ray temperature (keV) from White \etal (1997)
\\(5) - Classical cooling rates (M$_{\odot}$ yr$^{-1}$) from White \etal (1997)
\\$^a$ - No available \emph{Chandra} data
\\$^b$ - No available \emph{Chandra}, \emph{GALEX} or \emph{XMM-OM} data
\em{}
\label{sample}
\end{table}

\subsection{H$\alpha$: MMTF}

The MMTF has a very narrow bandpass ($\sim$5--12\AA) which can be
tuned to any wavelength over $\sim$5000-9200\AA (Veilleux \etal
2010). Coupled with the exquisite image quality at Magellan and the
wide field of the Inamori-Magellan Areal Camera \& Spectrograph
(IMACS), this instrument is ideal for detecting emission-line
filaments in distant clusters. During 2008-09, we observed all 23
clusters at both H$\alpha$ ($\lambda$=6563\AA) and continuum ($\pm$
60\AA), for a total of 20 minutes each. If H$\alpha$ was detected in a
20-minute exposure, it was followed up for an additional 40 minutes at
each wavelength. The typical image quality for these exposures was
$0.6 \pm 0.2^{\prime\prime}$

These data were fully reduced using the MMTF data reduction
pipeline\footnote{http://www.astro.umd.edu/$\sim$veilleux/mmtf/datared.html},
which performs bias subtraction, flat fielding, sky line removal,
cosmic ray removal, astrometric calibration and stacking of multiple
exposures (following Veilleux \etal 2010; see also Jones,
Bland-Hawthorn, Shopbell 2002). The continuum image was then PSF and
intensity matched to the narrow-band images to allow for careful
continuum subtraction. The stacked images were calibrated using
spectrophotometric standards from Oke (1990) and Hamuy \etal (1992,
1994) The error associated with our absolute photometric calibrations
is $\sim$15\%, which is typical for tunable filters and
spectrographs. Finally, the data were corrected for Galactic
extinction, following Cardelli \etal (1989) using reddening estimates
from Schlegel \etal (1998). We do not attempt to correct for intrinsic
extinction since the dust content of the optical filaments is not well
known. All of these procedures are described in detail in Veilleux
\etal (2010).

For systems with complicated morphologies, H$\alpha$ fluxes were
measured by creating (by eye) a region which generously traced the
H$\alpha$ emission and calculating the total signal within this
region. For more symmetric morphologies, a circular aperture centered
on the emission peak was used, with the radius chosen to contain all
of the obvious emission. Our measured H$\alpha$ fluxes (see Table A.1)
agree well with those in the literature (e.g., Cowie \etal 1983, Owen
\etal 1995, Edge 2001, Jaffe \etal 2005) for the 12 systems with
previously published data. In 9 systems we measure slightly higher
($\sim$ 1.7$\times$) total flux, most likely due to our better
sensitivity and larger field-of-view.  In all other cases, the
published ``H$\alpha$'' flux also include a contribution from [N~II]
$\lambda\lambda$ 6548, 6583 and is therefore overestimated. In all
cases, the H$\alpha$ emission measured from our data is not
contaminated by [N~II] emission.

\subsection{Near-UV: \emph{GALEX} and \emph{XMM-OM}}
One possible source of H$\alpha$ ionization in these clusters is
photoionization from hot, young stars. In this scenario we would
expect to see associated UV emission, perhaps with filamentary
morphology as well. In order to properly assess if this is the case,
we have used archival \emph{GALEX} (1350-1750\AA, 1750-2800\AA) and
\emph{XMM-OM} (1820-2320\AA, 2078-2517\AA, 2595-3215\AA) data for most
of the clusters in our sample.  The \emph{GALEX} data for Abell~1795
comes from our own program (GALEX GI Cycle 5 Program 31; PI
Veilleux). These data were corrected for Galactic extinction,
following Cardelli \etal (1989) using Galactic reddening estimates
from Schlegel \etal (1998).

\begin{figure}[htbp]
\begin{center}
\includegraphics[width=0.75\textwidth]{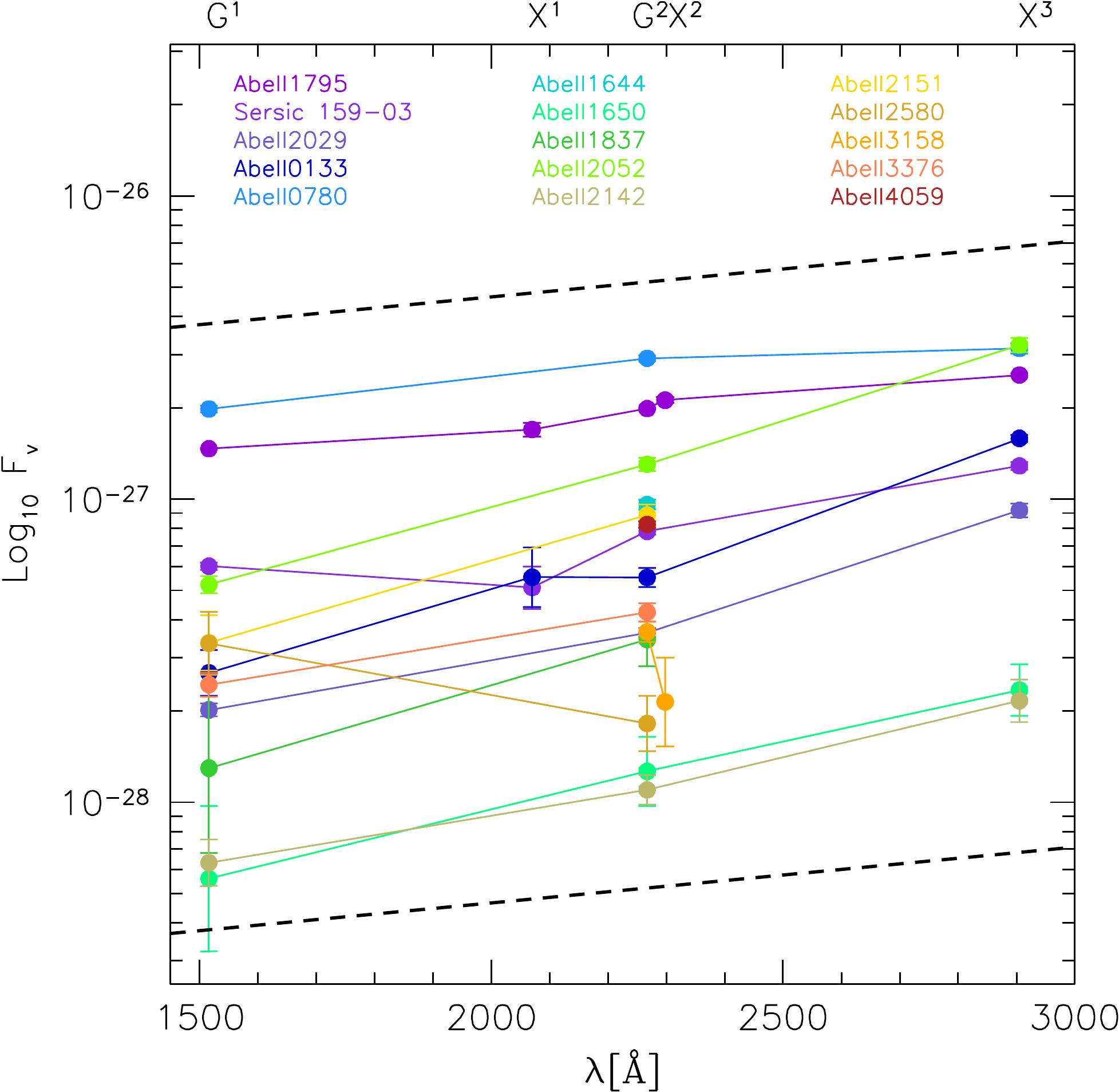}
\caption{UV SEDs for 15 clusters in our sample with observations at
multiple wavelengths. All fluxes have been corrected for Galactic
extinction. The two \emph{GALEX} filters are denoted by G$^1$ and
G$^2$, while the three \emph{XMM-OM} filters are denoted by X$^1$,
X$^2$ and X$^3$. The dashed lines represent the mean SED slope. These
SEDs are sufficiently flat to justify merging our 18 GALEX NUV
(2267\AA) and 3 XMM-OM UVW1 (2905\AA) observations into a single
near-UV sample.}
\label{uvsed}
\end{center}
\end{figure}

In order to determine the star formation rate (SFR) from the NUV flux,
we follow the prescription described in Kennicutt (1998). This
technique assumes that SFR $\propto$ $F_{\nu}$ with no dependence on
wavelength ($F_{\nu}$ is flat) over the range of 1500-2800\AA. In
Figure \ref{uvsed}, we confirm that this assumption is roughly valid
for the clusters in our sample and, thus, if dust is present it does
not follow the galactic reddening law. For 18 clusters, we have
\emph{GALEX} near-UV (2267\AA) data, while the remaining 3 have
\emph{XMM-OM} UVW1 (2905\AA) data. The difference in flux that we
predict between these filters (from Figure \ref{uvsed}) is
$\sim1.3\times$. Thus, we proceed with merging these two sources of UV
data with the caveat that UV fluxes for Abell 0478, 0496 and 1650 may
be $\sim$30\% too high.

\begin{figure}[htbp]
\begin{center}
\begin{tabular}{cc}
\includegraphics[width=0.45\textwidth]{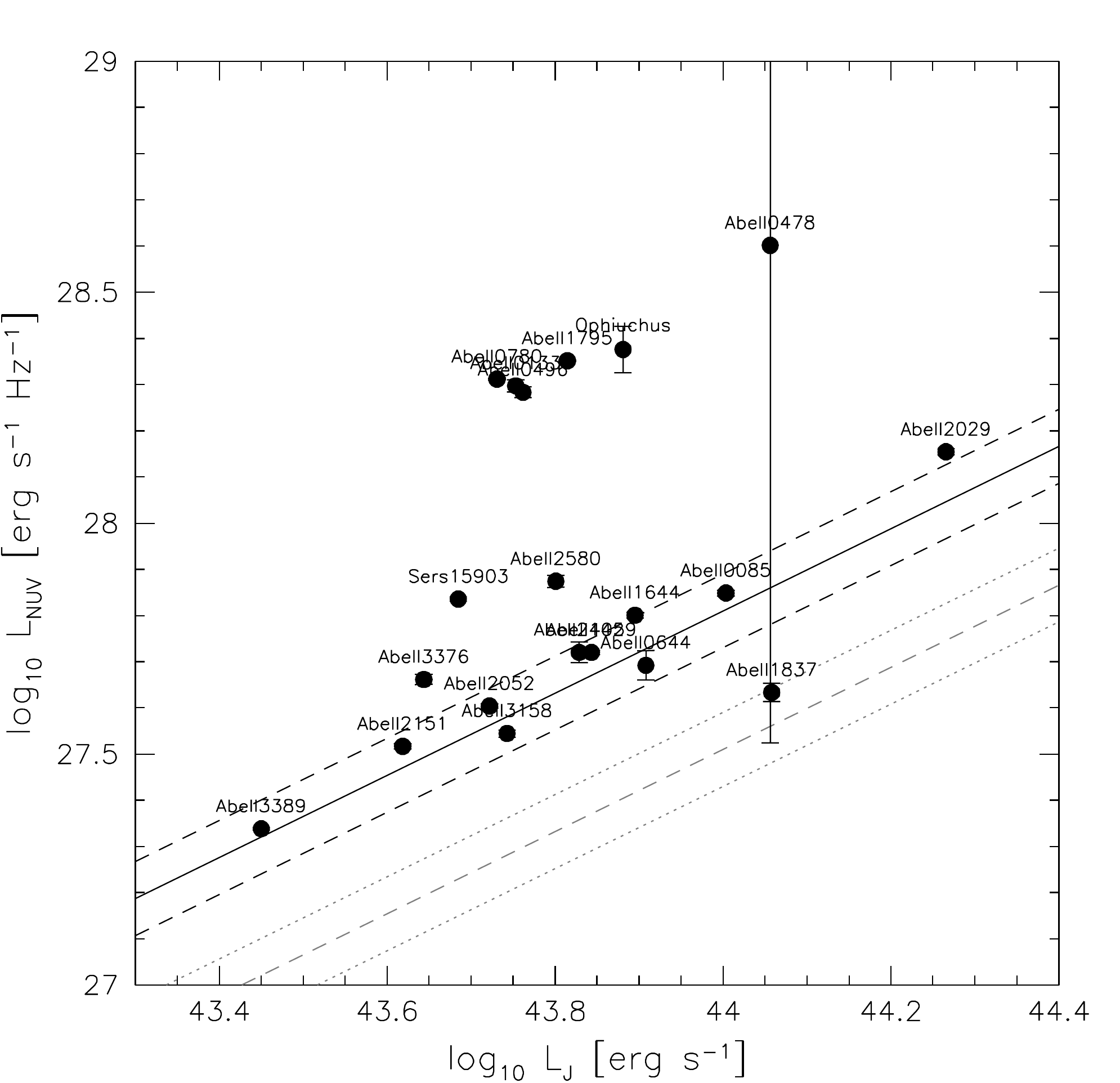} & 
\includegraphics[width=0.45\textwidth]{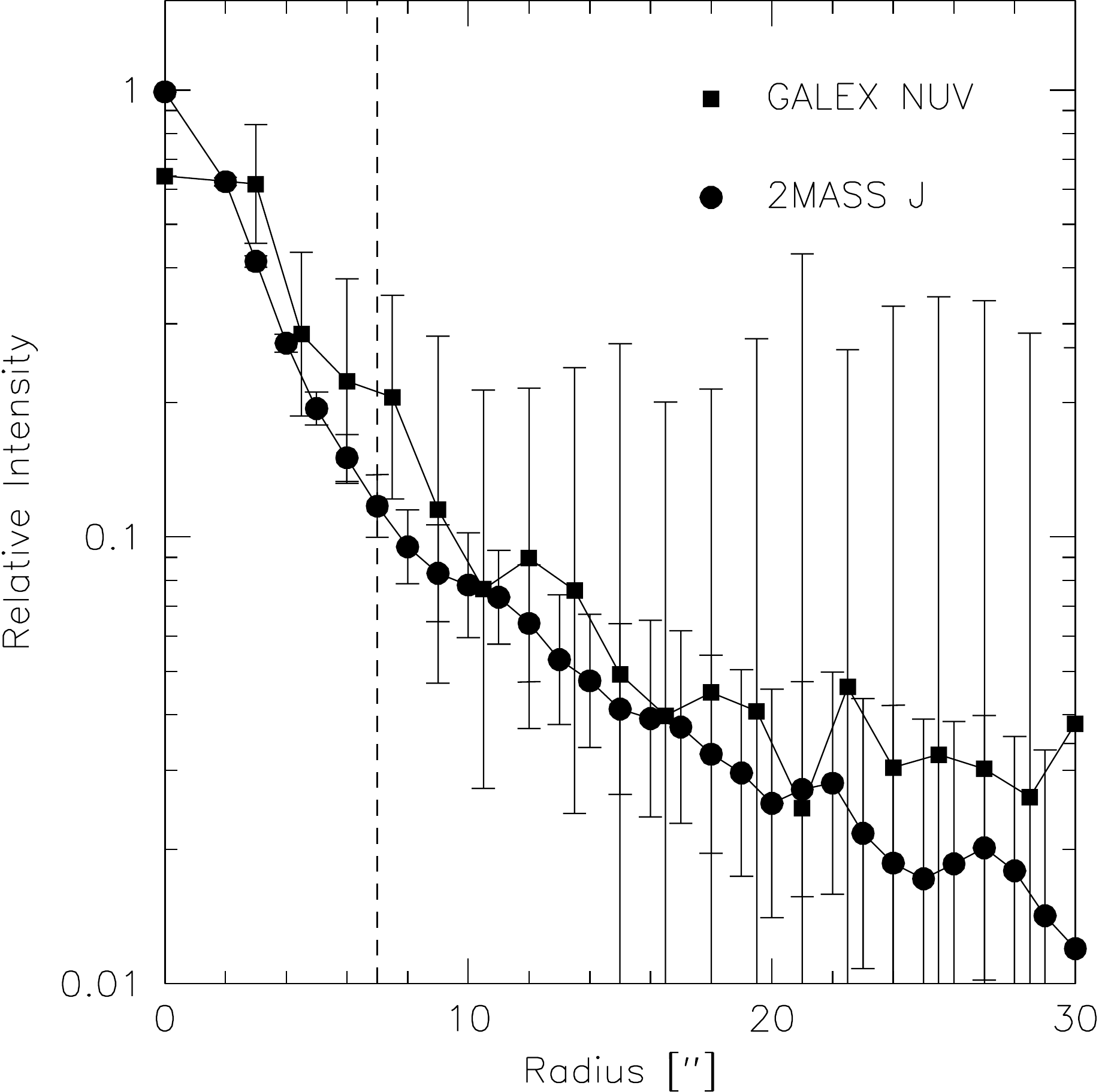} \\
\end{tabular}
\caption{Left panel: GALEX NUV (1750 -- 2800 \AA) versus J-band
luminosity for the 21 BCGs in our sample with archival UV data. The
errorbars on L$_J$ are smaller than the point size in general. The
gray dashed line represents the scaling relation determined by Hicks
\etal (2010) for a 7$^{\prime\prime}$ aperture, with the scatter in
this relation represented by the gray dotted line.  The black, solid
line is our estimate of the contribution to the total NUV luminosity
from old stellar populations for a 30$^{\prime\prime}$ aperture. This
line was chosen to fit the several points with the lowest NUV/J
ratios, with the assumption that these BCGs have negligible ongoing
star formation. The dashed lines represent the scatter in this
relation determined by Hicks \etal (2010). The zero-point difference
between this work and Hicks+ is due to the size of the aperture
considered. By removing this NUV contribution from old stellar
populations, we can consider only the excess due to young stars. Right
panel: Radial intensity profile of the BCG in Abell~3158 (a passive,
non-star forming BCG) at NUV and J-band. An offset has been applied to
the NUV intensity to match the J-band luminosity inside of
7$^{\prime\prime}$. Beyond this radius, denoted by a vertical dashed
line, the NUV intensity is consistently higher. Thus, by increasing
the aperture to 30$^{\prime\prime}$, the NUV/J ratio should increase.}
\label{excess}
\end{center}
\end{figure}

Accompanying J-band images were also obtained from 2MASS in order to
remove the expected NUV emission from the old stellar population.
Motivated by the techniques described in Hicks \etal (2005, 2010), we
attempt to remove the NUV contribution from old stellar populations, which we infer
from the J-band data. The relationship between the GALEX NUV and 2MASS J-band fluxes follow a powerlaw: 
\begin{equation}
\log_{10} L_{NUV} =  C_1 + C_2 \log_{10} L_J
\end{equation}
Hicks \etal define a non-star forming control sample in order to
calibrate this relationship, consisting of 17 cluster ellipticals and
22 BCGs in clusters without measurable cooling flows. The BCGs in this
control sample cover a similar range in J-band luminosity and redshift
to the clusters in our sample. Hicks \etal (2010) find $C_1$ = -0.85
$\pm$ 0.016 and $C_2$ = 0.89 $\pm$ 0.045, with a scatter of 0.08,
based on fluxes measured in 7$^{\prime\prime}$ apertures. Since we are
interested in detecting extended emission, we instead use a
30$^{\prime\prime}$ aperture, which has the net effect of changing the
zero-point of this relation (due to the NUV-J color gradient). Thus,
we adjust the zero-point by hand so that it passes through the points
with the lowest NUV/J ratios. We find $C_1$ = -0.55 and note that the
observed scatter about this fit matches well that quoted by Hicks
\etal (2010). Fig. \ref{excess} shows this fit, along with that found
by Hicks \etal (2010), for the BCGs in our sample. Using this
relation, we scaled the measured J-band fluxes to obtain the NUV
contribution from the old stellar population, and then subtracted this
from the measured NUV flux. Our value of $C_1$ is slightly larger
than that of Hicks \etal (2010), due to our use of a larger
aperture. We show in the right-hand panel of Fig. \ref{excess} that,
beyond the 7$^{\prime\prime}$ aperture used by Hicks, the NUV/J ratio
is higher at all radii out to 30$^{\prime\prime}$. Thus, a slight
increase in $C_1$ is justified.

We do not attempt to correct for intrinsic extinction, due to the fact
that not all of our clusters have reddening estimates, thus the lower
envelope in Fig. \ref{excess} could also contain dusty, star forming
galaxies. However, from Crawford \etal (1999), we find that the
typical intrinsic extinction in BCGs with strong H$\alpha$ emission is
E(B-V) = 0.29 $\pm$ 0.135. This would induce a scatter of 0.5 in the
NUV-J relation. Since the measured scatter is much lower (0.08), we
propose that the majority of the galaxies used in the Hicks \etal
(2010) NUV-J relation are relatively dust-free. Furthermore, it is
unclear how the dust is distributed in the BCG (e.g. Lauer \etal 2005)
and, thus, any intrinsic extinction correction would require a
significant amount of guesswork.

\subsection{X-Ray: \emph{Chandra}}
Archival data from the \emph{Chandra X-ray Observatory} were retrieved
for 19 of our 23 sample clusters.  These data were reprocessed with
CIAO (version 4.1.2) and CALDB (version 4.1.1) using the latest time-dependent gain adjustments and maps to create new level 2 event
files. Due to the large angular extent of the clusters in our sample,
we were required to construct blank-sky background event files, using
the ACIS blank-sky background database, to properly account for
background flux. The new level 2 event files were cleaned for flares,
using the $lc\_clean$ routine, by examining the light-curve and
removing any spurious bursts in intensity. These data cleaning and
calibration procedures are all outlined in detail in the CIAO science
threads\footnote{http://cxc.harvard.edu/ciao/threads/}.

In order to separate any filaments or interesting morphology from the
X-ray halo, we performed an unsharp masking technique on each image,
subtracting a 10$^{\prime\prime}$ Gaussian smoothed image from a
1.5$^{\prime\prime}$ Gaussian smoothed image. The resulting image highlights
any fine structure in the X-ray morphology.

For each cluster, background-subtracted spectra were extracted using
$dmextract$. Updated response files were created using $mkacisrmf$ and
$mkwarf$, following the CIAO science threads. Counts were grouped into
bins with 20 counts per bin, over the range 0.3 to 11.0 keV. Spectra
were extracted in a variety of regions to better
understand the relationship between the ICM and the H$\alpha $
emission. These regions are described below, followed by a description
of the spectral modeling which we apply to these spectra.

\subsubsection{Radial Profiles}
For all clusters, spectra were extracted from a series of
concentric annuli. The maximum annulus was chosen by eye (typically
the largest allowable in the CXO field-of-view) and then the interior
annuli were chosen iteratively (following Sun \etal 2009) with
r$_{out}$/r$_{in}$ = 1.25--1.6 so that each annulus contained a
minimum of 10,000 counts. This choice of spacing allows for a reliable
deprojection and fit to the spectra. The innermost annulus was chosen
to have an inner radius of 2 pixels, to exclude any contribution from
the AGN. The data in this inner region of 2-pixel radius is used to
identify possible AGN emission.

\subsubsection{Filaments}
\begin{figure}[htb]
\begin{center}
\includegraphics[width=0.8\textwidth]{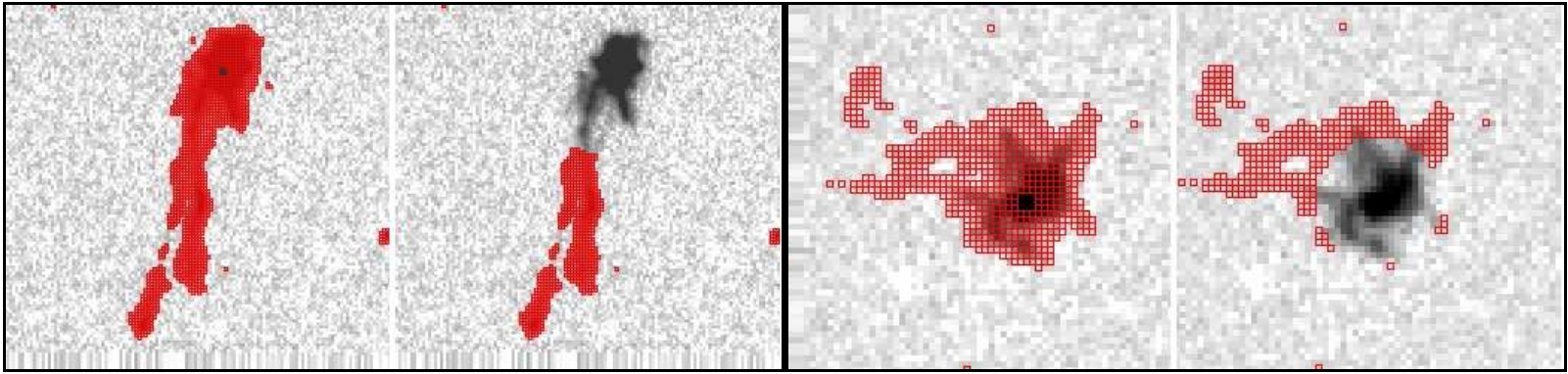}
\caption{Examples of the on-filament extraction area (shown in red)
for Abell 1795 (left) and Abell 0496 (right). The left-most frame in
each pair covers the full radial range, while the right-most covers
the range 0.3R$_{H\alpha}$ $<$ r $<$ R$_{H\alpha}$. The off-filament
extraction area contains all unmarked pixels in the same radial
range.}
\label{fil_example}
\end{center}
\end{figure}

In order to properly investigate any possible connection between the
ICM and the presence of H$\alpha$ filaments, we also extract X-ray
spectra in regions where we detect H$\alpha$ emission. A mask was
created from the MMTF images by first registering them to the CXO
images using the IRAF \emph{wregister} task. The typical error
involved in this process is dependent on the relative astrometric
calibrations, which was typically a fraction of a pixel, or $\sim$
0.1$^{\prime\prime}$. The registered image was then smoothed by
1$^{\prime\prime}$ using a Gaussian kernel. Finally, all pixels with a
flux greater than 3-$\sigma$ above the background and within
R$_{H\alpha}$ (defined as the outermost radius of H$\alpha$ emission,
measured from the peak of the H$\alpha$ emission, see \S3.1) were
included in the mask. We extracted spectra in four different regions:
(1) 0 $<$ r $<$ R$_{H\alpha}$, coincident with H$\alpha$, (2) 0 $<$ r
$<$ R$_{H\alpha}$, anti-coincident with H$\alpha$, (3)
0.3R$_{H\alpha}$ $<$ R $<$ R$_{H\alpha}$, coincident with H$\alpha$,
(4) 0.3R$_{H\alpha}$ $<$ R $<$ R$_{H\alpha}$, anti-coincident with
H$\alpha$.  These spectra allow for a direct comparison between X-ray
derived properties inside and outside of the H$\alpha$ filaments, both
near the center and at larger radii. An example of the on-filament
extraction regions is shown in Figure \ref{fil_example}, for Abell
1795 and Abell 0496.

\subsubsection{Spectral Modeling}
In order to derive physical quantities from the X-ray spectra, we used
the XSPEC spectral fitting package (Arnaud 1996).  As mentioned above,
spectra in radial annuli have also been extracted. This allows for the
computation of various parameters such as $T_X$ and $n_e$ as a
function of radius. Since the emission in any given annulus is a
combination of emission at that radius and also at larger radii seen
in projection, we need to deproject the data in each radial bin. We do
this using two different techniques: direct spectral deprojection
(DSDEPROJ; Russel \etal 2008) and the PROJCT routine in XSPEC. We
find DSDEPROJ to yield more stable radial profiles, in agreement with
Russel \etal (2008), who showed that PROJCT produces highly fluctuating
radial profiles when applied to a spectrum that represents gas at
multiple temperatures. Thus the majority of the
analysis uses the results from this deprojection. However, it should
be mentioned that all of the relevant quantities which we derive from
these profiles are robust to the spectral deprojection method
used. Along with the spectral deprojection, we used a photoelectric
absorption model (PHABS) multiplied by a model for thermal
bremsstrahlung emission from a hot diffuse gas (MEKAL). The elements
were assumed to be present in the Solar ratios, as measured by Anders
\& Grevesse (1989). The free parameters for this combination of models
are the column density of hydrogen, $N_H$, the plasma temperature in
keV, $T_X$, the metal abundance of the plasma, $Z$, and the
normalization of the model, which is directly related to the electron
density in the plasma, $n_e$.  The model normalization ($N_{MEKAL}$)
is related to the electron density by:
\begin{equation}
N_{MEKAL} ={ {10^{-14}}\over{4\pi [D_A(1+z)^2]}}\int n_en_HdV
\end{equation}
where $D_A$ is the angular diameter distance to the source, in cm, and
$dV$ is the volume of the emitting region. We assume that the
X-ray halo is spherical, in order to make the volume calculation
straightforward. In order to determine the electron density inside and
outside of the ionized filaments, we also need to assume
something about their thickness. We discuss this problem further
in a later section.

Finally, in order to determine the mass deposition rate in the cooling
flow (\.{M}$_{spec}$), we also apply a second set of models to the
radially-binned spectra. This new set of models consists of the same
deprojection and photoelectric absorption mentioned above, multiplied
by the combination of a thermal plasma (MEKAL) and a second plasma
component that is cooling (MKCFLOW). The MKCFLOW model has two
temperature parameters, high T and low T, which we lock to the
temperature of the plasma and to 0.1 keV, respectively. The choice of
the low temperature is arbitrary - we simply need it to be lower than
the Chandra detection limit ($\sim$~0.5~keV) so that this represents
gas that has cooled out of the X-ray regime. The metal abundance in
the cooling gas is also required to be the same as the plasma. Thus,
the only free parameter in this new component is the normalization,
which is precisely the spectrally-determined cooling flow rate,
\.{M}$_{spec}$.

\subsection{1.4 GHz Radio: NVSS}
The total 1.4 GHz radio flux at the center of a cluster is a useful
diagnostic of AGN activity. For the 20 clusters with $\delta >
-40^{\circ}$ in our sample, we used the 1.4 GHz fluxes measured from
the NRAO VLA Sky Survey (NVSS; Condon \etal 1998).

\section{Results}
\subsection{Warm Ionized Filaments}

This survey is the largest sample of cooling flow clusters
with high-resolution, narrow-band imaging to date. The success of this
survey has been striking, with multiple extended filaments seen in
8/23 clusters and slightly extended or nuclear emission seen in an
additional 7/23 clusters. The remaining 8 clusters have no detectable
H$\alpha$ emission. This detection rate of $\sim$ 65\% is
consistent with the value of 71$^{+9}_{-14}$ found by Edwards \etal
(2007) for BCGs in cooling flow clusters (for BCGs in both cooling and
non-cooling clusters, Edwards \etal find the fraction to be $\sim$
15\%). Figure \ref{bigfig} shows the unsharp masked X-ray and
archival UV data, along with the new MMTF continuum and H$\alpha$
images, centered on the BCG. Upon inspection of these images, the
following trends are immediately evident:
\begin{itemize}
\item There is a strong correlation between the X-ray and
H$\alpha$ morphology. Clusters with ionized filaments tend to have
structure in the X-ray. This structure is brightest at lower energies
(0.5-2 keV, see Figure \ref{xray-rgb}), suggesting that the optical filaments
live in or near cooler X-ray structures.
\item There does not appear to be a correlation between the presence
of filaments and the presence of other galaxies near the BCG. In fact,
Abell~2151 and Abell~2029 each have one or more significant companions
within the optical radius of the BCG but exhibit no detectable
H$\alpha$ emission. Of the 7 clusters with extended H$\alpha$
filaments, only 3 (Abell~0478, Abell~1644, Abell~2052) have
significant companions within the optical radius of the BCG. In a select few
cases, the direction of the detected filaments may correlate with the
position of nearby galaxies (e.g., Abell~0478, Abell~1644).
\item The observed warm filaments are never wider than the MMTF PSF
($\sim 0.6^{\prime\prime}$), except in the case of Sersic 159-03. At a
typical redshift of 0.06 (d $\sim$ 250 Mpc), this corresponds to an
upper limit on the typical filament width of 0.7 kpc.
\item In some cases, where the filaments are very extended
(Abell~1644, Abell~1795, Sersic~159-03), there is evidence for
matching morphology in the near-UV. This is especially true in the
case of Abell~1795, which we have confirmed with much higher
resolution near-UV images in MV09 and McDonald \etal (in prep;
hereafter M+10). However, it is challenging to match the NUV and
H$\alpha$ morphologies in general due to the mismatch in the PSF
widths.
\end{itemize}

The properties of the observed H$\alpha$ filaments are summarized in
Table~A.1 and a detailed discussion of individual clusters can be
found in the appendix. Based on the properties of the H$\alpha$
emission, the sample can be divided into 3 categories: those with
complex, extended filaments (type I), those with nuclear emission or
only slightly extended (i.e. a few PSF widths) emission (type II) and
those with no observable emission (type III). We use this convention
throughout the remainder of our discussion to easily identify trends
at other wavelengths which could explain both the flux and morphology
of the H$\alpha$ emission.

\begin{figure*}[p]
\centering
\begin{tabular}{c}
Abell 0085\\
\includegraphics[width=0.9\textwidth]{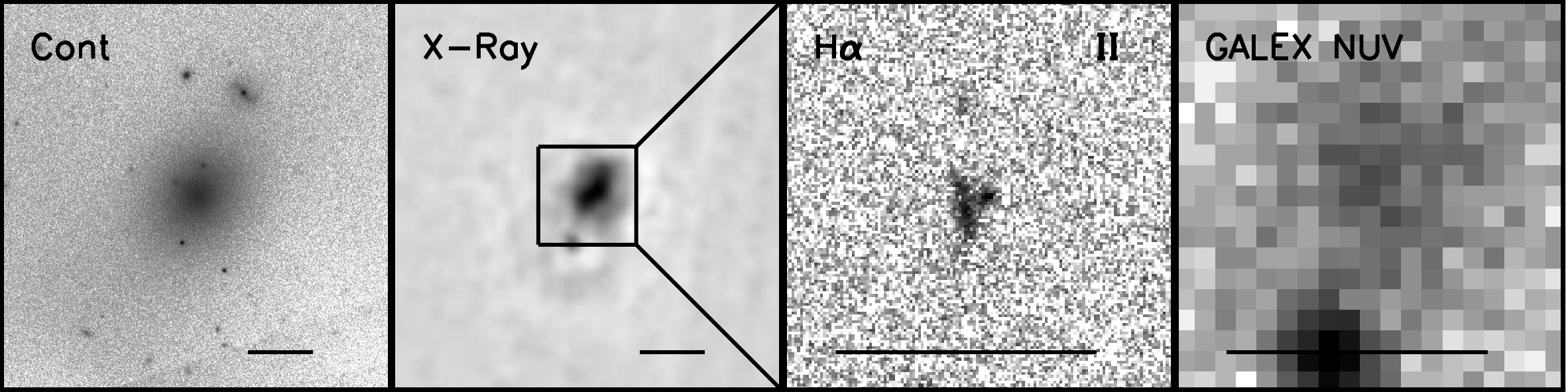} \\
Abell 0133\\
\includegraphics[width=0.9\textwidth]{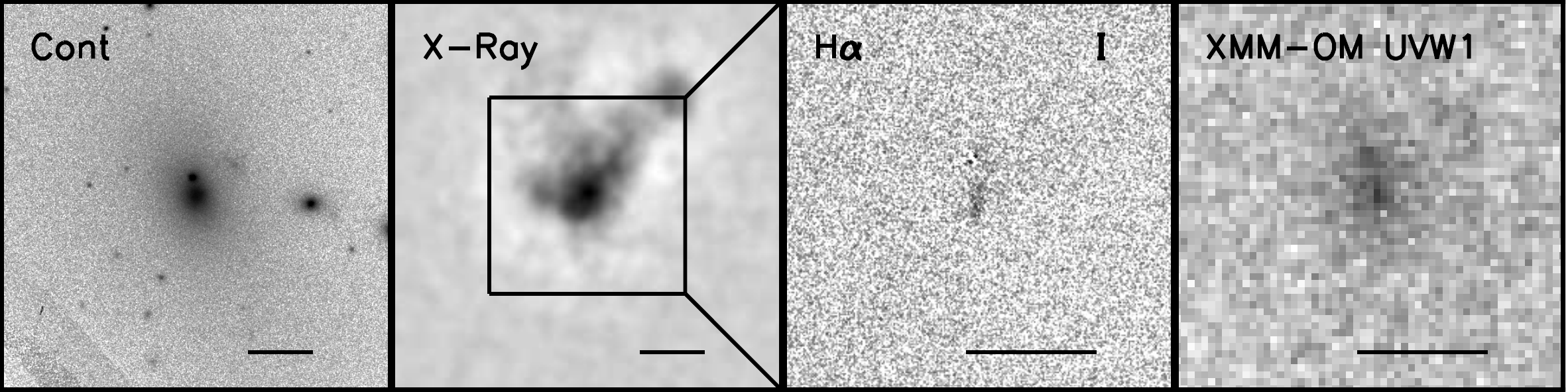} \\
Abell 0478\\
\includegraphics[width=0.9\textwidth]{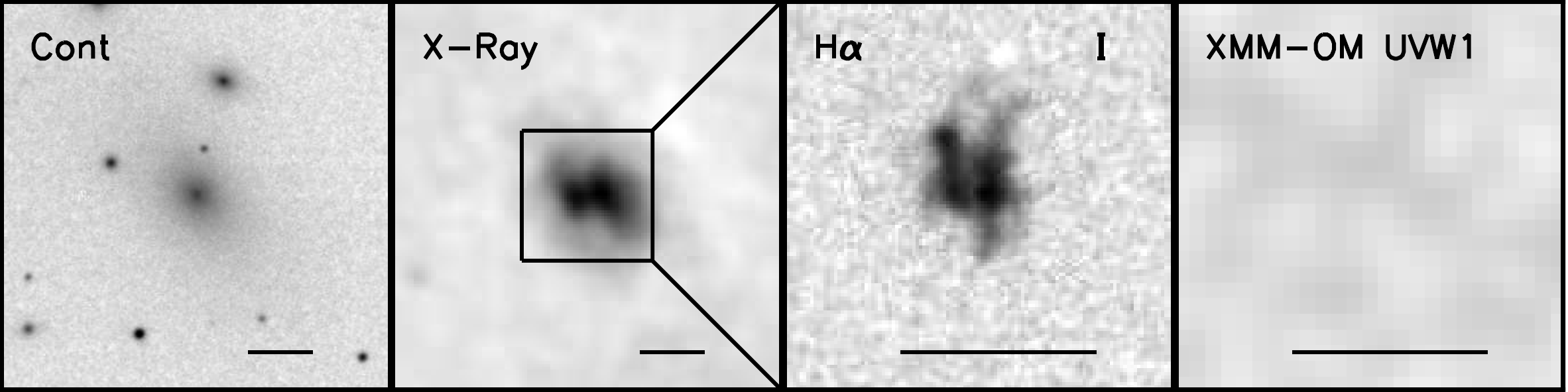} \\
Abell 0496\\
\includegraphics[width=0.9\textwidth]{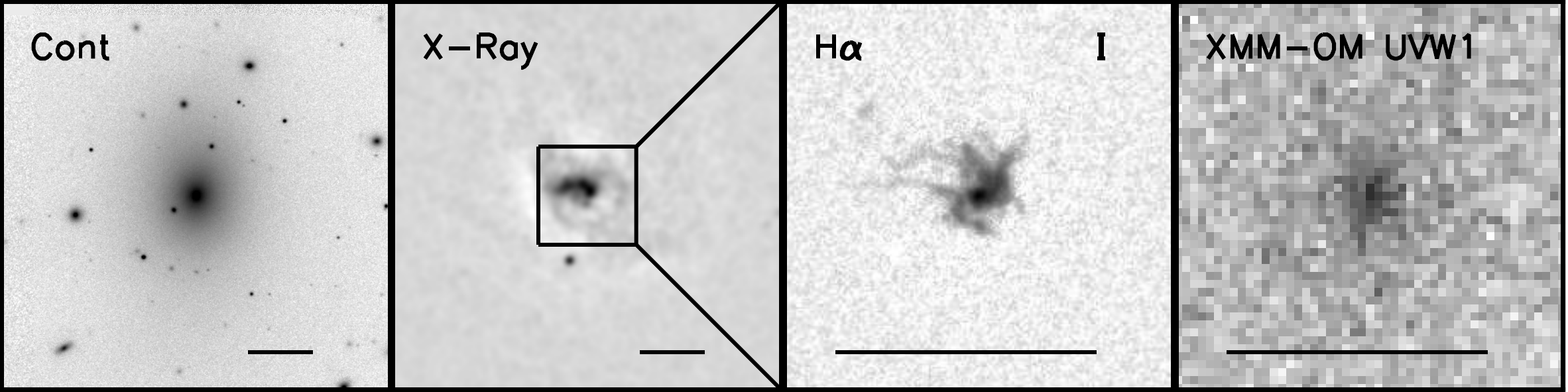} \\
\end{tabular}
\caption{Multi-wavelength data for the 21 clusters in our sample with
NUV (GALEX, XMM-OM) and H$\alpha$ (MMTF) data. From left to right the
panels are: 1) MMTF red continuum image, 2) Unsharp masked CXO X-ray
image, 3) MMTF continuum-subtracted H$\alpha$ image, 4) GALEX/XMM-OM
NUV image. The horizontal scale bar in all panels represents 20
kpc. The X-ray and red continuum images are on the same scale, and the
H$\alpha$ and NUV data are on the same zoomed-in scale. The square
region in the X-ray panels represents the field of view for the
zoomed-in H$\alpha$ and NUV panels.}
\label{bigfig}
\end{figure*}

\addtocounter{figure}{-1}

\begin{figure*}[p]
\centering
\begin{tabular}{c}
Abell 0644\\
\includegraphics[width=0.9\textwidth]{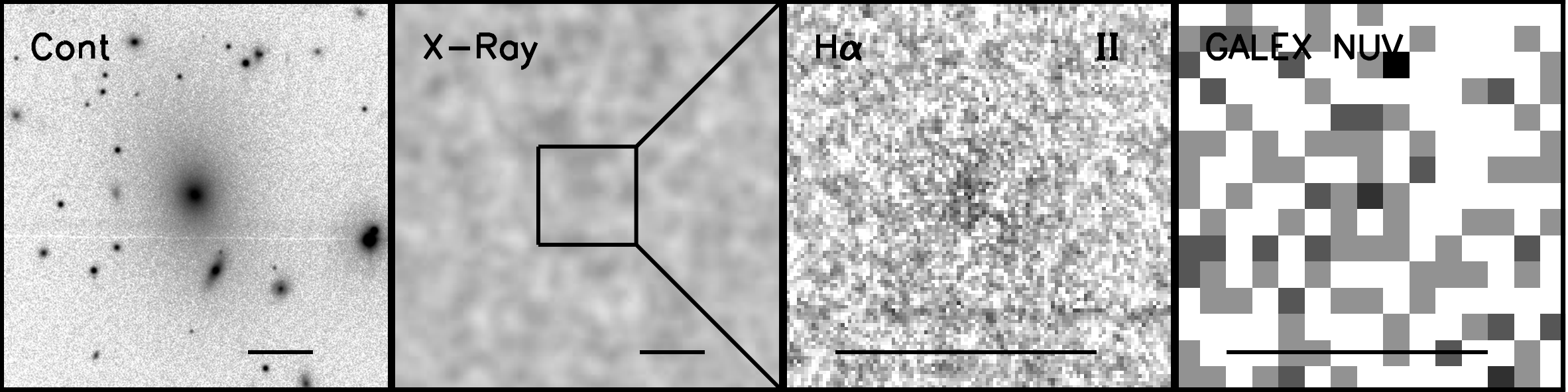} \\
Abell 0780 (Hydra A)\\
\includegraphics[width=0.9\textwidth]{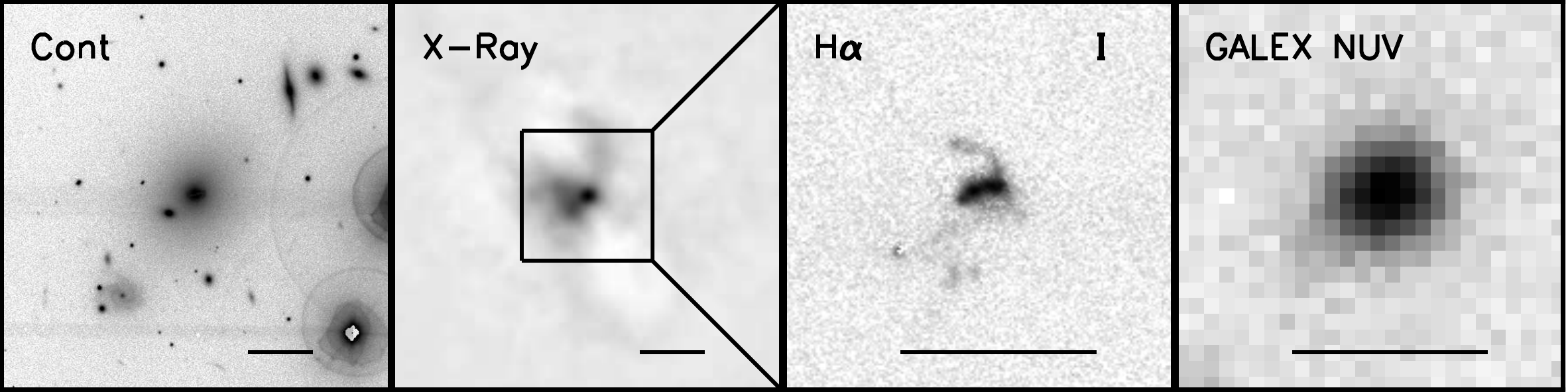} \\
Abell 1644\\
\includegraphics[width=0.9\textwidth]{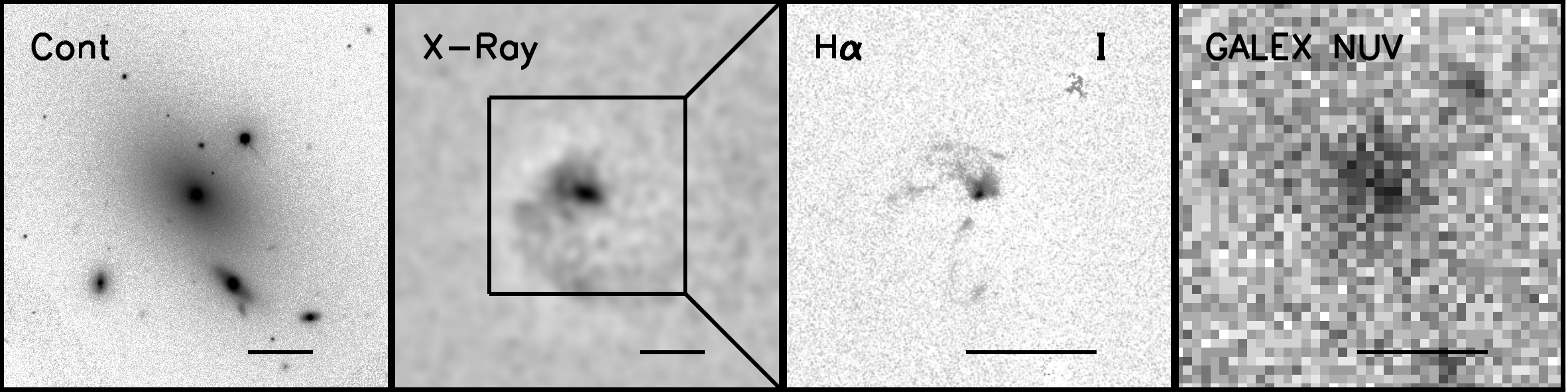} \\
Abell 1650\\
\includegraphics[width=0.9\textwidth]{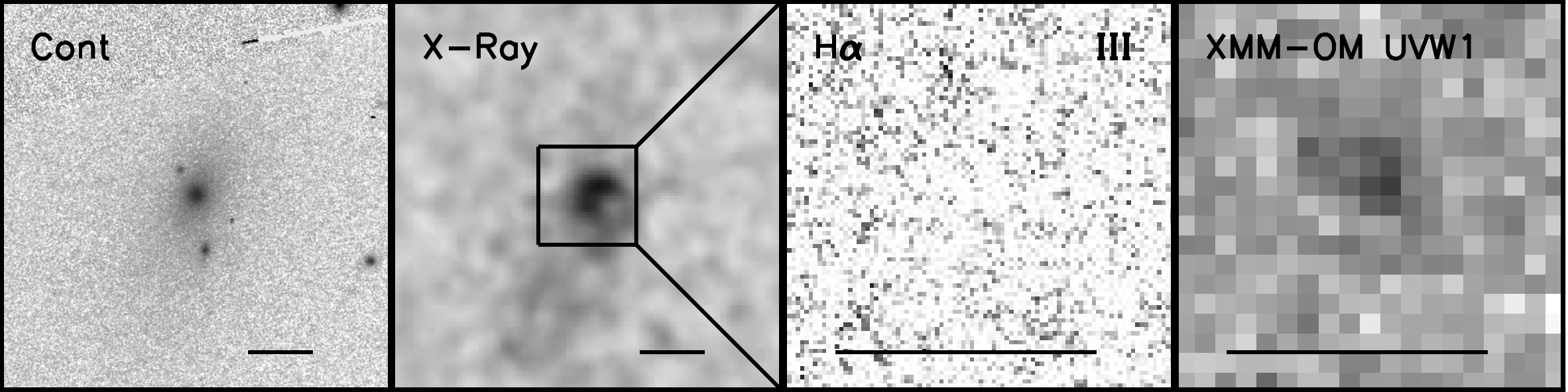} \\
Abell 1795\\
\includegraphics[width=0.9\textwidth]{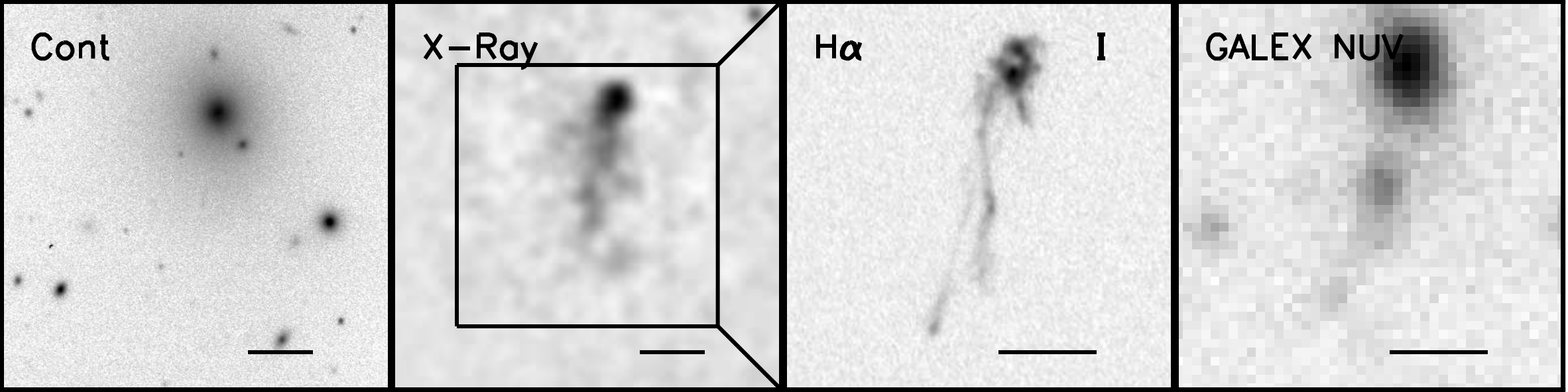} \\
\end{tabular}
\caption{Continued.}
\end{figure*}

\addtocounter{figure}{-1}

\begin{figure*}[p]
\centering
\begin{tabular}{c}
Abell 1837\\
\includegraphics[width=0.9\textwidth]{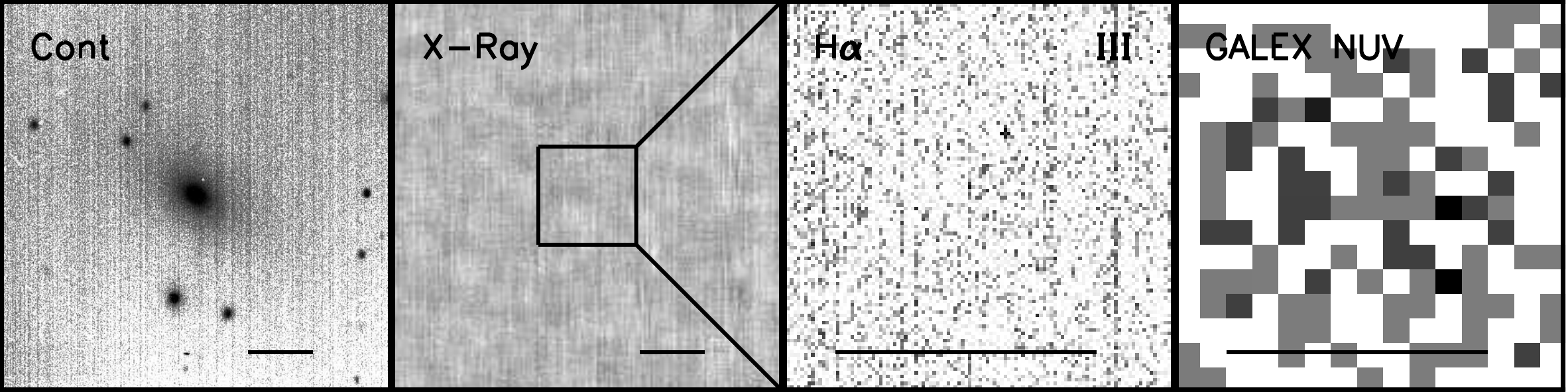} \\
Abell 2029\\
\includegraphics[width=0.9\textwidth]{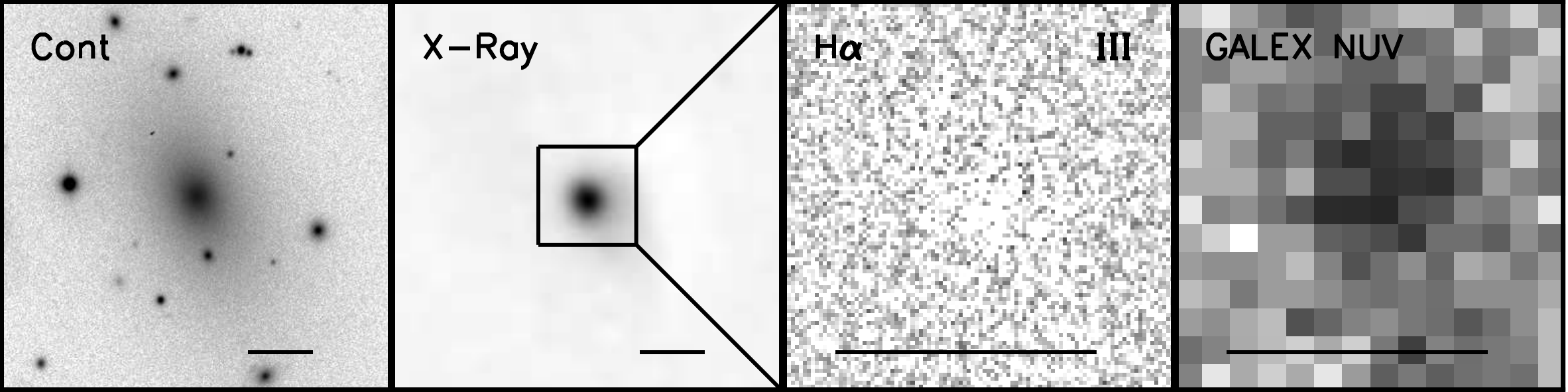} \\
Abell 2052\\
\includegraphics[width=0.9\textwidth]{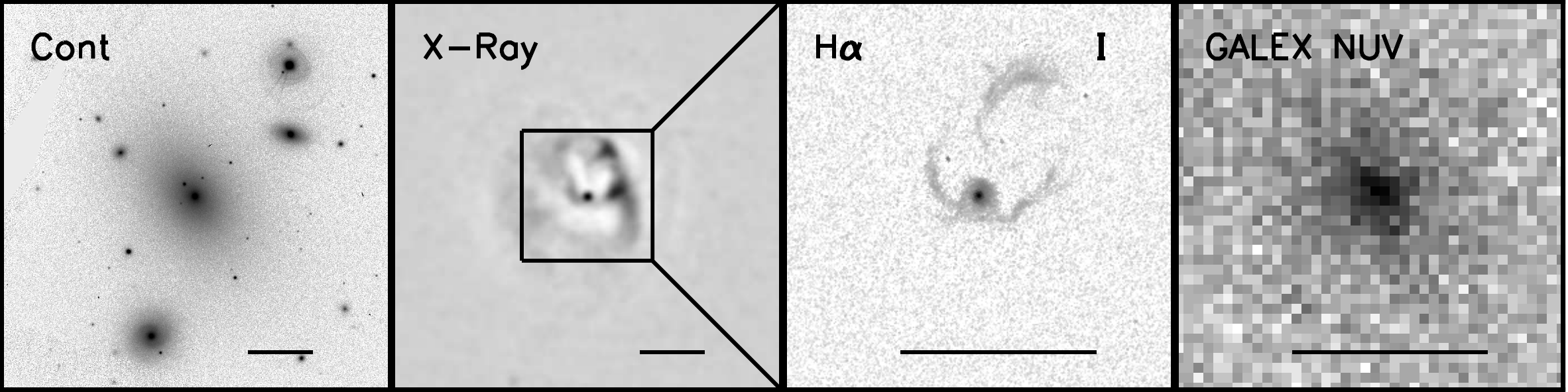} \\
Abell 2142\\
\includegraphics[width=0.9\textwidth]{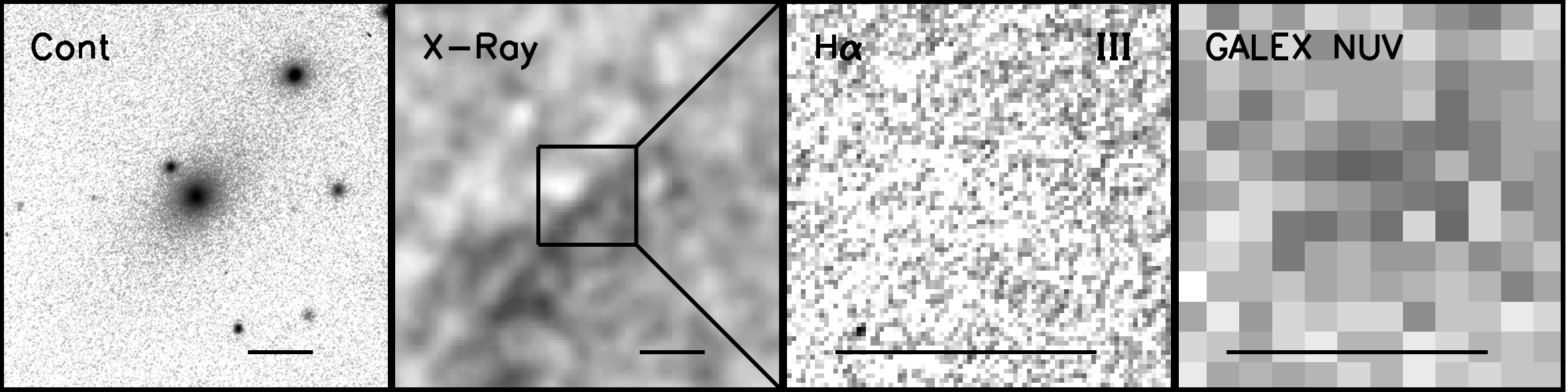} \\
Abell 2151\\
\includegraphics[width=0.9\textwidth]{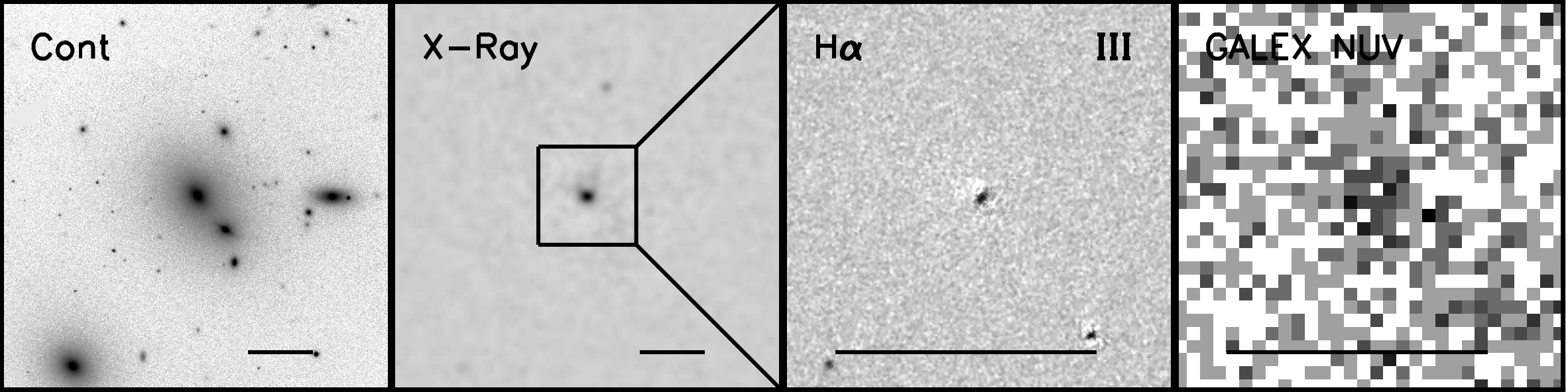} \\
\end{tabular}
\caption{Continued.}
\end{figure*}

\addtocounter{figure}{-1}

\begin{figure*}[p]
\centering
\begin{tabular}{c}
Abell 2580\\
\includegraphics[width=0.9\textwidth]{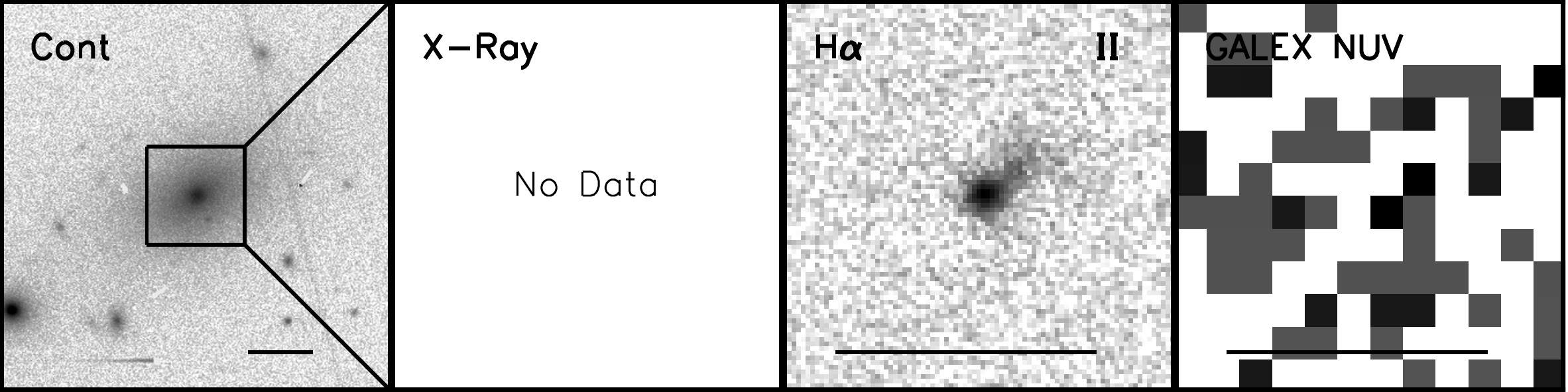} \\
Abell 3158\\
\includegraphics[width=0.9\textwidth]{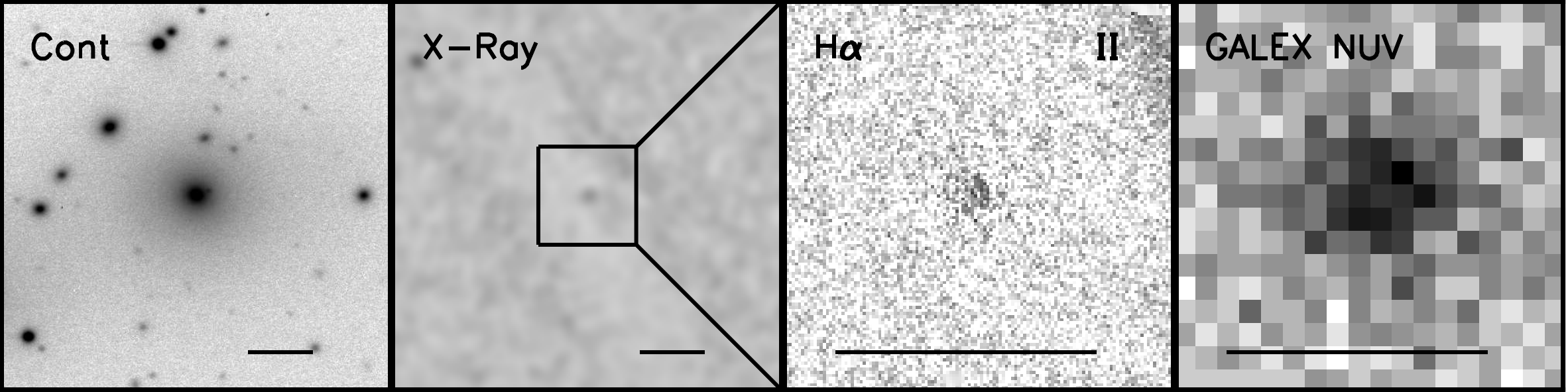} \\
Abell 3376\\
\includegraphics[width=0.9\textwidth]{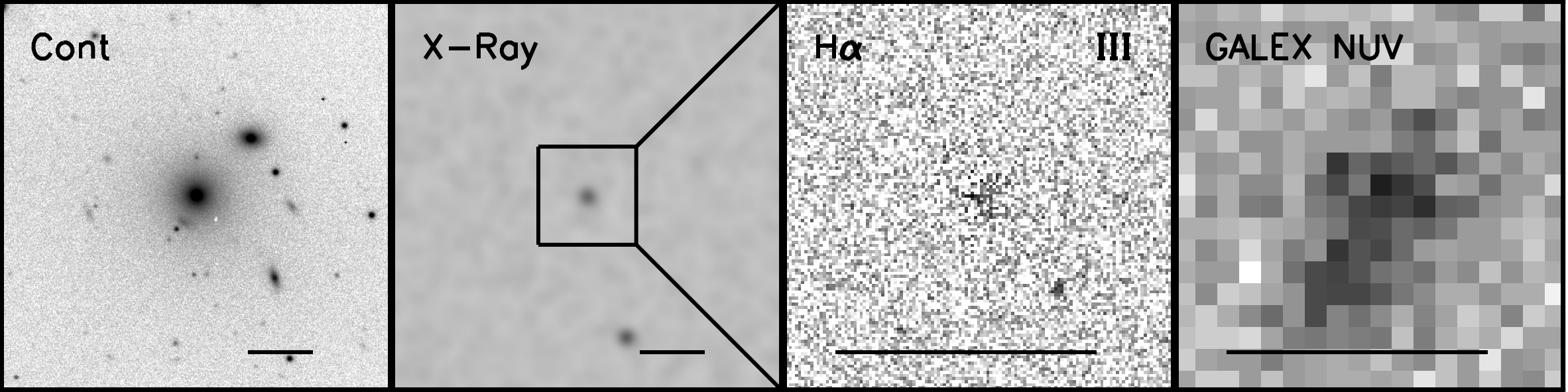} \\
Abell 3389\\
\includegraphics[width=0.9\textwidth]{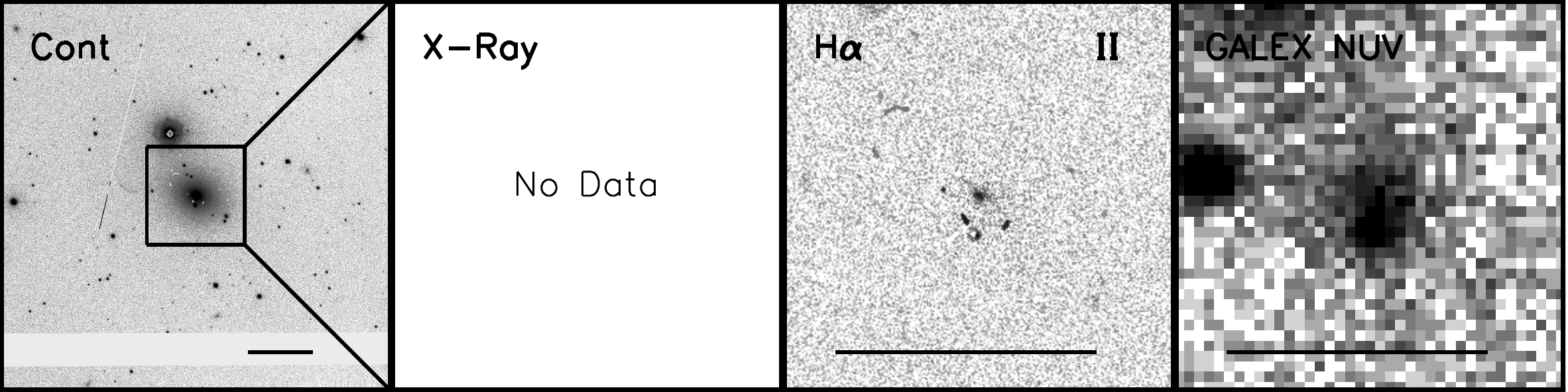} \\
Abell 4059\\
\includegraphics[width=0.9\textwidth]{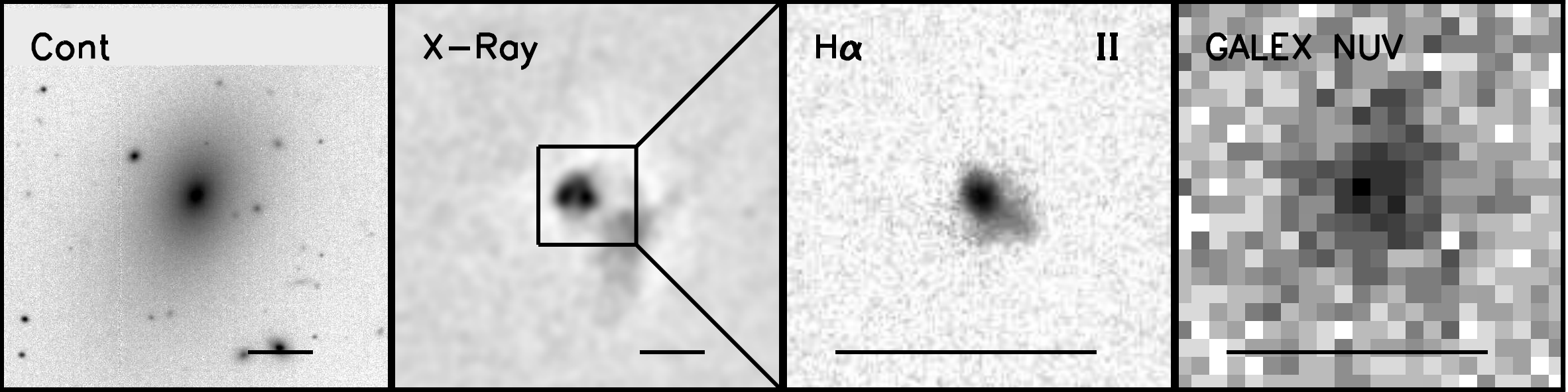} \\
\end{tabular}
\caption{Continued.}
\end{figure*}

\addtocounter{figure}{-1}

\begin{figure*}[htbp]
\centering
\begin{tabular}{c}
Ophiuchus\\
\includegraphics[width=0.9\textwidth]{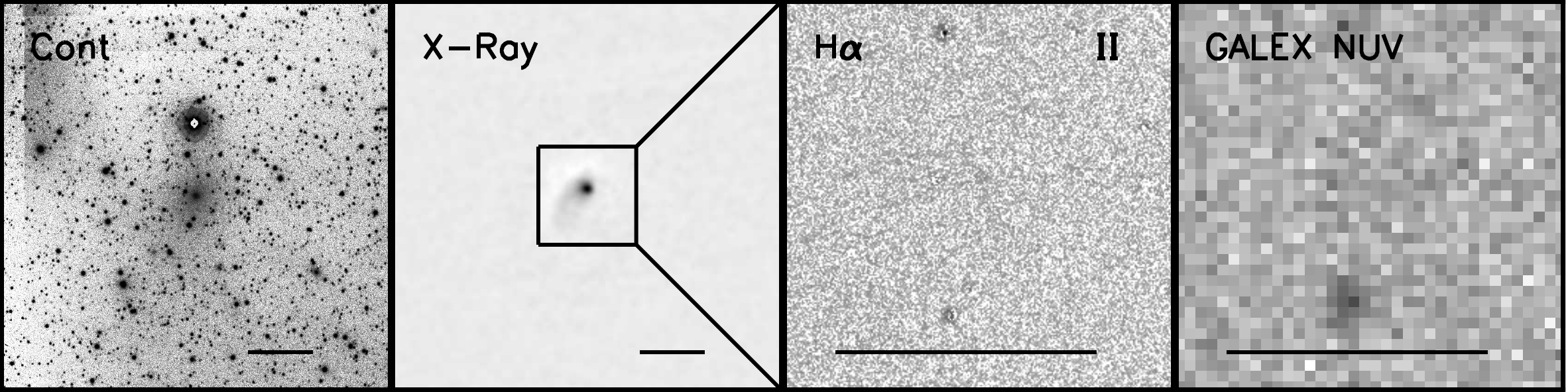} \\
Sersic 159-03\\
\includegraphics[width=0.9\textwidth]{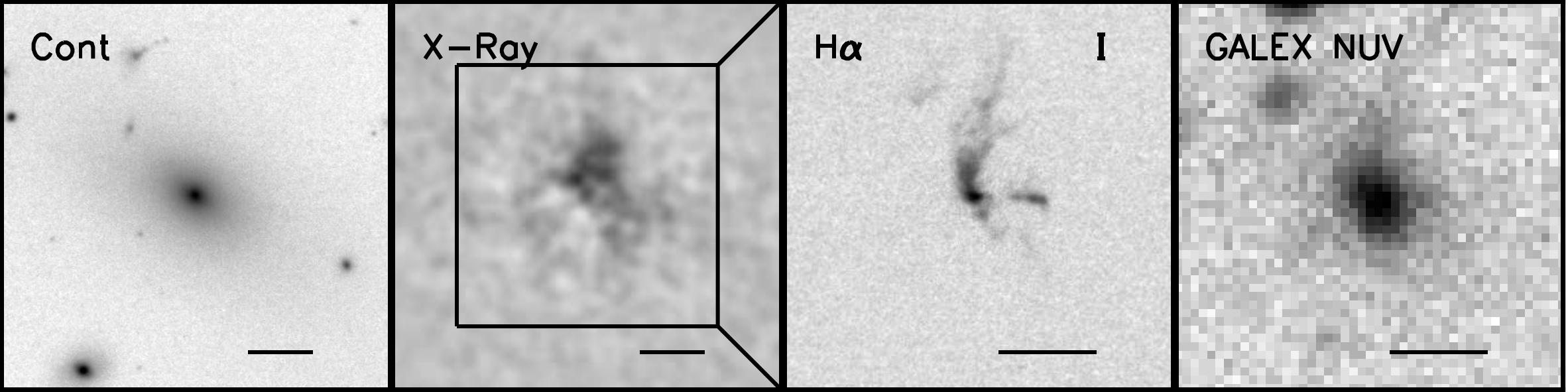} \\
\end{tabular}
\caption{Continued.}
\end{figure*}


\subsection{Star Formation}
A likely explanation for the presence of H$\alpha$ emission is
photoionization by massive, young stars. However, while several
studies to date have found star formation in the central cluster
galaxy, very little work has been done on correlating the observed
H$\alpha$ filaments with young stars for a large sample of
clusters. As mentioned earlier, archival near-UV data from
\emph{GALEX} and \emph{XMM-OM} exist for 21 of our 23 sample
clusters. Figure \ref{uvha} shows that there is a weak
correlation (Pearson R = 0.51) between the near-UV and H$\alpha$
luminosities. This result is not surprising since the cooling flow in
these systems is allegedly depositing some cold gas onto the BCG and
triggering some star formation activity. However, there is
considerably more scatter than one would expect if the ionization was
purely from continuous star formation, suggesting that modifications
to the star formation law (i.e. altering the IMF), or alternative
physical mechanisms entirely, may be needed to explain the observed
NUV/H$\alpha$ ratios.

\begin{figure*}[htb]
\centering
\begin{tabular}{c c c}
\includegraphics[width=0.3\textwidth,height=0.3\textwidth,angle=0]{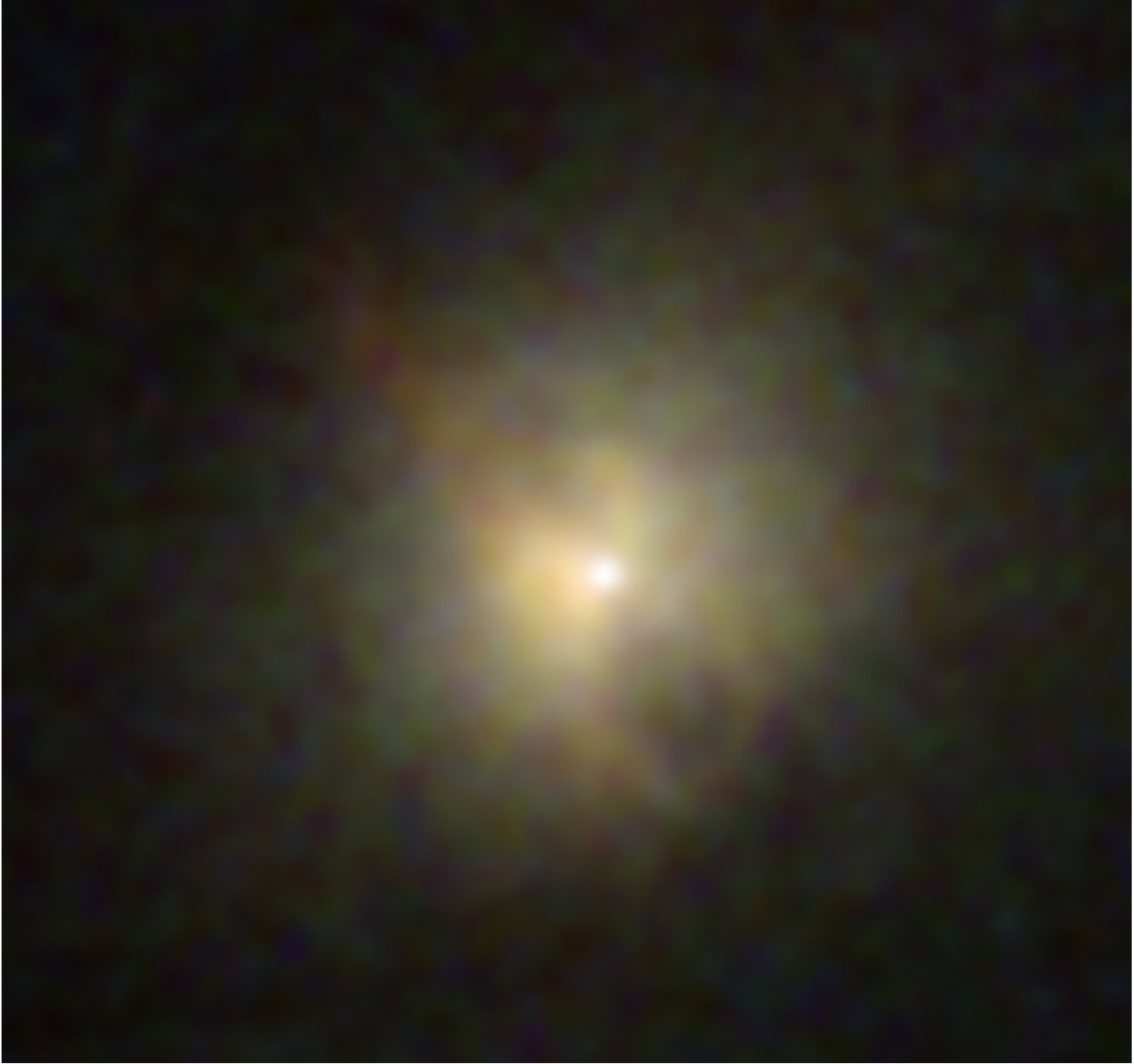} &
\includegraphics[width=0.3\textwidth,height=0.3\textwidth,angle=0]{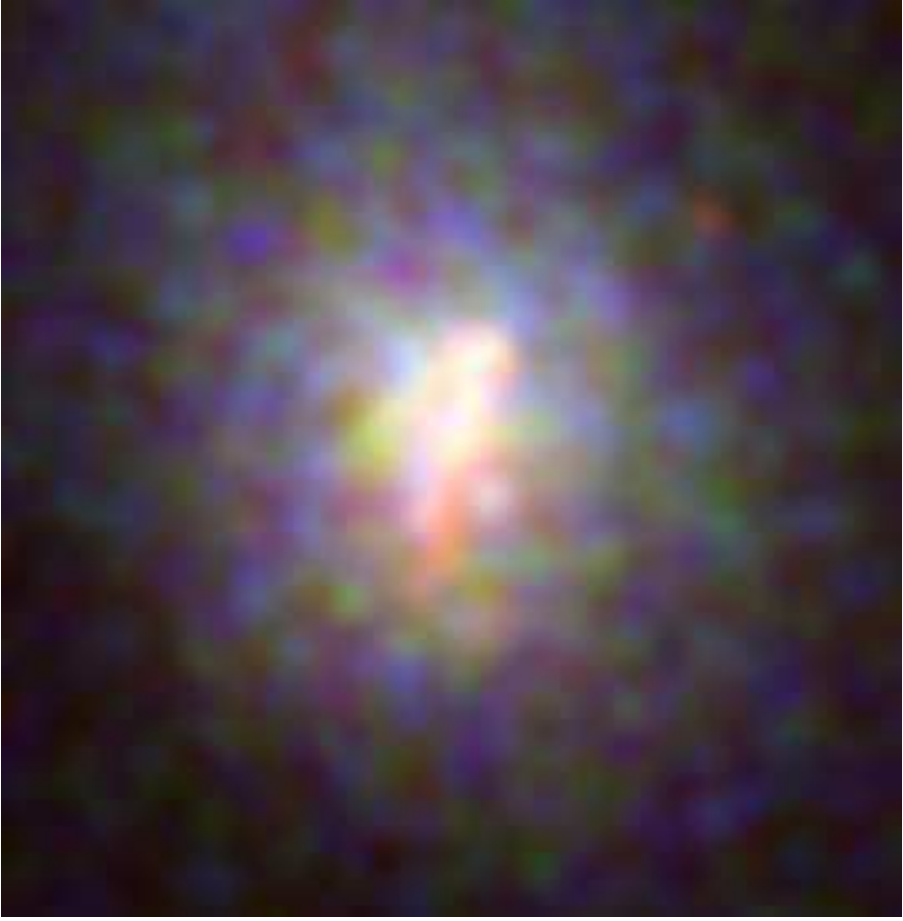} &
\includegraphics[width=0.3\textwidth,height=0.3\textwidth,angle=0]{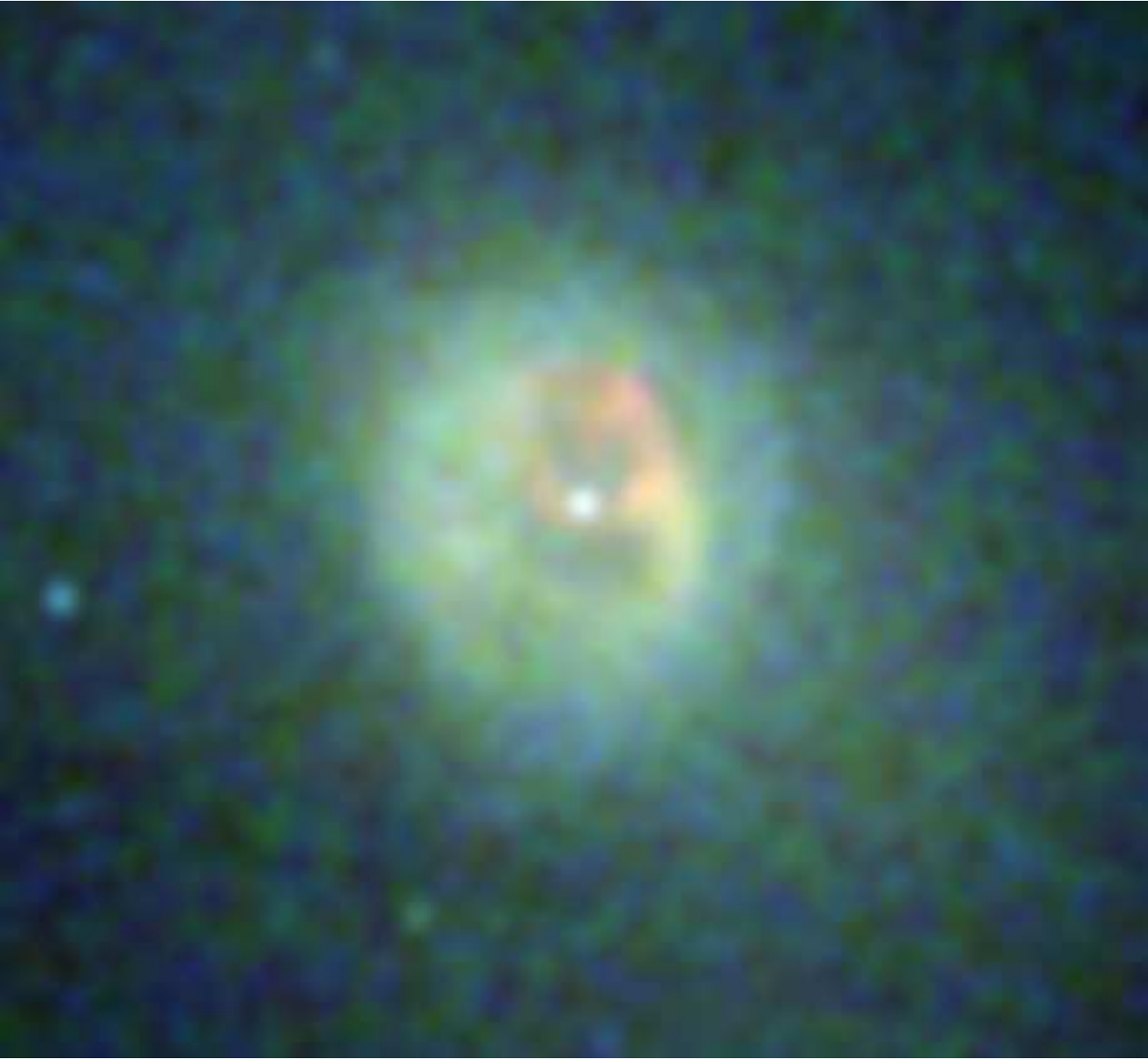} 
\end{tabular}
\caption{False color X-ray images of three clusters in our
sample. From left to right, these are Abell~0780 (Hydra A), Abell~1795
and Abell~2052. The colors are 0.5-2 keV (red), 2-4 keV (green) and
4-8 keV (blue). The asymmetry in the X-ray ICM is clearly associated
with the coolest (0.5-2 keV) gas.}
\label{xray-rgb}
\end{figure*}

\begin{figure}[p]
\begin{center}
\includegraphics[width=0.9\textwidth]{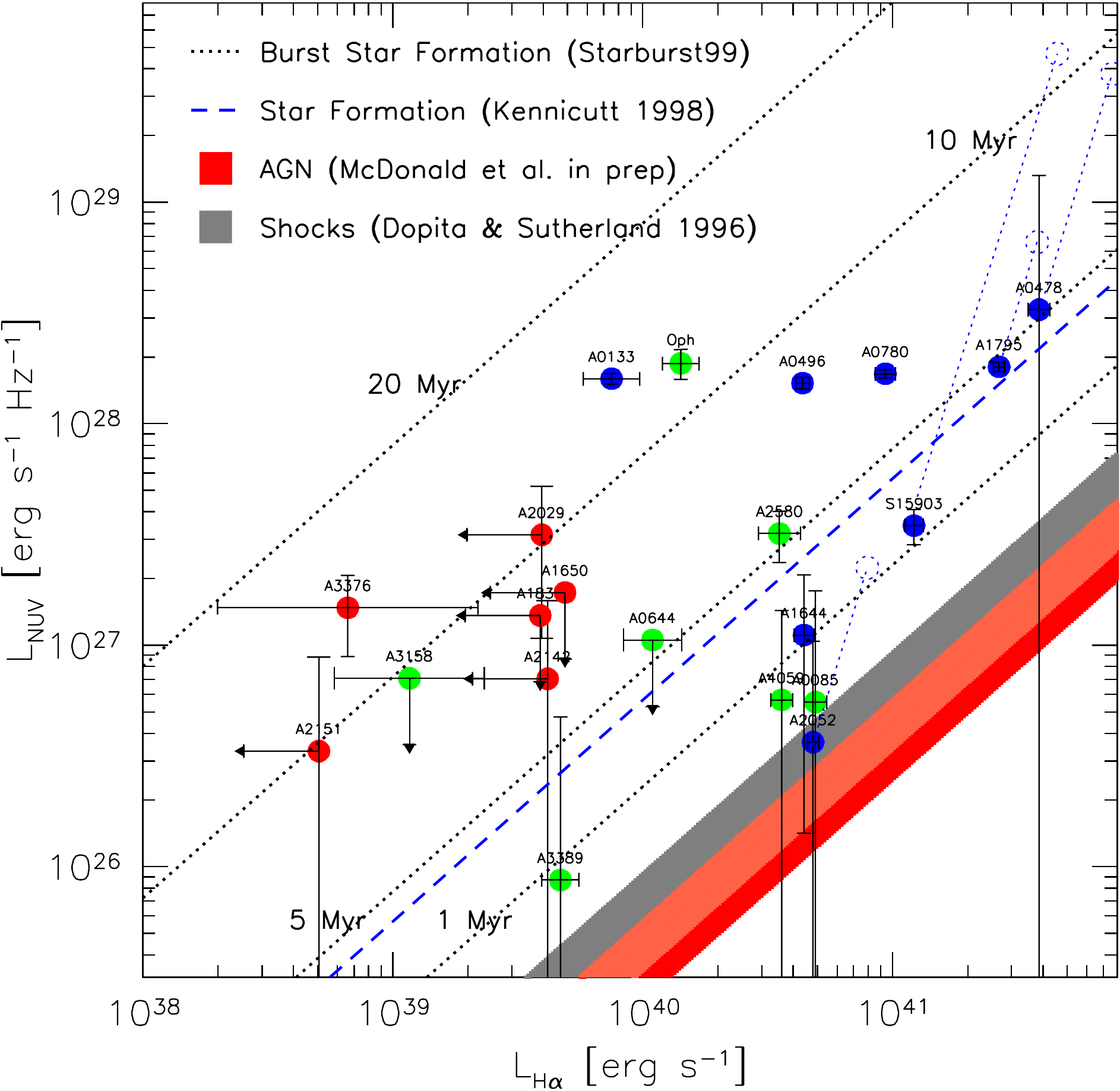}
\caption{Global GALEX NUV versus MMTF H$\alpha$ luminosities for the
21 clusters in our sample with archival GALEX data. The dashed blue
line represents ongoing star formation, described by Kennicutt
(1998). The dotted black lines describe the evolution of the
NUV/H$\alpha$ ratio as a starburst ages from 1--20 Myr (Starburst99;
Leitherer \etal 1999). The shaded grey area shows the range of
NUV/H$\alpha$ ratios expected from shocks (Dopita \& Sutherland 1996),
while the shaded red area shows the expected NUV/H$\alpha$ ratio for
the AGN, based on HST FUV imaging of cluster cores (M+10 in prep). The
orange area represents the overlap between these two regimes. The
point color refers to the classification described in \S3.1: type I
(blue), type II (green) and type III (red). The dotted lines/circles show the effect of correcting
for intrinsic extinction, with reddening estimates taken from Crawford \etal (1999) and 
Jaffe \etal (2005). The large deviations from
ongoing star formation (Kennicutt 1998), suggest that non-stellar
processes may be responsible for some of the observed H$\alpha$
emission. Contamination from the nucleus likely plays an important
role in the observed scatter.}
\label{uvha}
\end{center}
\end{figure}

In Figure \ref{uvha}, we also show the NUV/H$\alpha$ ratios for
several different processes, including continuous star formation
(described by Kennicutt 1998).  Using models from Starburst99
(Leitherer \etal 1999) we age a burst of star formation from 1--20 Myr
in order to see the evolution of the NUV and H$\alpha$ flux. These
models assume a metallicity of 0.4Z$_{\odot}$ and a Kroupa IMF (Kroupa
2001) with $\alpha = 3.3$. The NUV/H$\alpha$ ratio is quite sensitive
to the IMF assumed, but the overall trends remain the same. As the
massive stars die off, the UV/H$\alpha$ ratio increases rapidly.  This
scenario is promising, since several clusters have NUV/H$\alpha$
ratios above the value predicted by Kennicutt et al. (1998) and there
are very few physical mechanisms that can produce such ratios. In this
scenario, star formation took place in a burst as gas crosses
R$_{cool}$ and the NUV/H$\alpha$ ratio is a direct indicator of the
time elapsed since this burst. If this is indeed the reason for the
high UV/H$\alpha$ ratios, we would expect higher resolution UV data to
show age gradients along the length of any filaments. This trend was
not seen in Abell~1795 (McDonald \& Veilleux 2009), although there was
a slight offset between the UV and H$\alpha$ emission which made
quantifying a gradient difficult. An alternative explanation for the
high NUV/H$\alpha$ ratios is the dependence of this quantity as a
function of H$\alpha$ surface density (Meurer \etal 2009). For star
formation in regions with low H$\alpha$ surface brightness
(i.e. UGCA44) the UV/H$\alpha$ ratio is a factor of $\sim$5$\times$
higher than for a region with higher surface brightness
(i.e. NGC~1566). Thus, both scenarios of an aging burst of star
formation and star formation in low surface density environments can
yield the scatter above the Kennicutt relation that we observe. A
third potential source of high UV/H$\alpha$ ratios is intrinsic
reddening due to dust in the BCG. We show, in Figure \ref{uvha}, the
effect of correcting for intrinsic reddening, using extinction
measurements from Crawford \etal (1999) and Jaffe \etal
(2005). Applying this correction tends to move points \emph{further}
above the Kennicutt relation, suggesting that the high UV/H$\alpha$
ratios we measure are not due to dust in the BCG.

Figure \ref{uvha} presents the \emph{global} NUV and H$\alpha$
luminosities, since the poor spatial resolution of the
\emph{GALEX}/\emph{XMM-OM} data does not allow us to discriminate
between filament and nuclear emission. In MV09, however, we show that
the UV/H$\alpha$ ratio in the nucleus of Abell~1795 is quite low
(L$_{FUV}$/L$_{H\alpha}$ $\sim$ 7$\times$10$^{-14}$~Hz$^{-1}$). In M+10 (in prep), we show that this is
typical for AGN (central point source) in BCGs based on HST far-UV
imaging of cooling flow cluster cores. This means that the effect of
including the nuclear contribution in the luminosity calculation would
move the data below the Kennicutt relation.

Finally, we show the UV/H$\alpha$ ratio expected from shock heating
(Dopita \& Sutherland 1996). Again, this produces low ratios and thus
could only be a strong contributer in very few of the clusters. Most
notably, Figure \ref{uvha} shows that the NUV/H$\alpha$ ratio for
Abell~2052 is consistent with shock heating, and we observe that the
H$\alpha$ emission traces the outer shell of a cavity carved in the
ICM by a radio jet seen at 1.4GHz (see Figure \ref{bigfig}). This is
strong evidence for the H$\alpha$-emitting gas to be shock heated
by an AGN outflow in this object.

\subsection{X-Ray Profiles}

\begin{figure}[p]
\begin{center}
\includegraphics[width=0.9\textwidth]{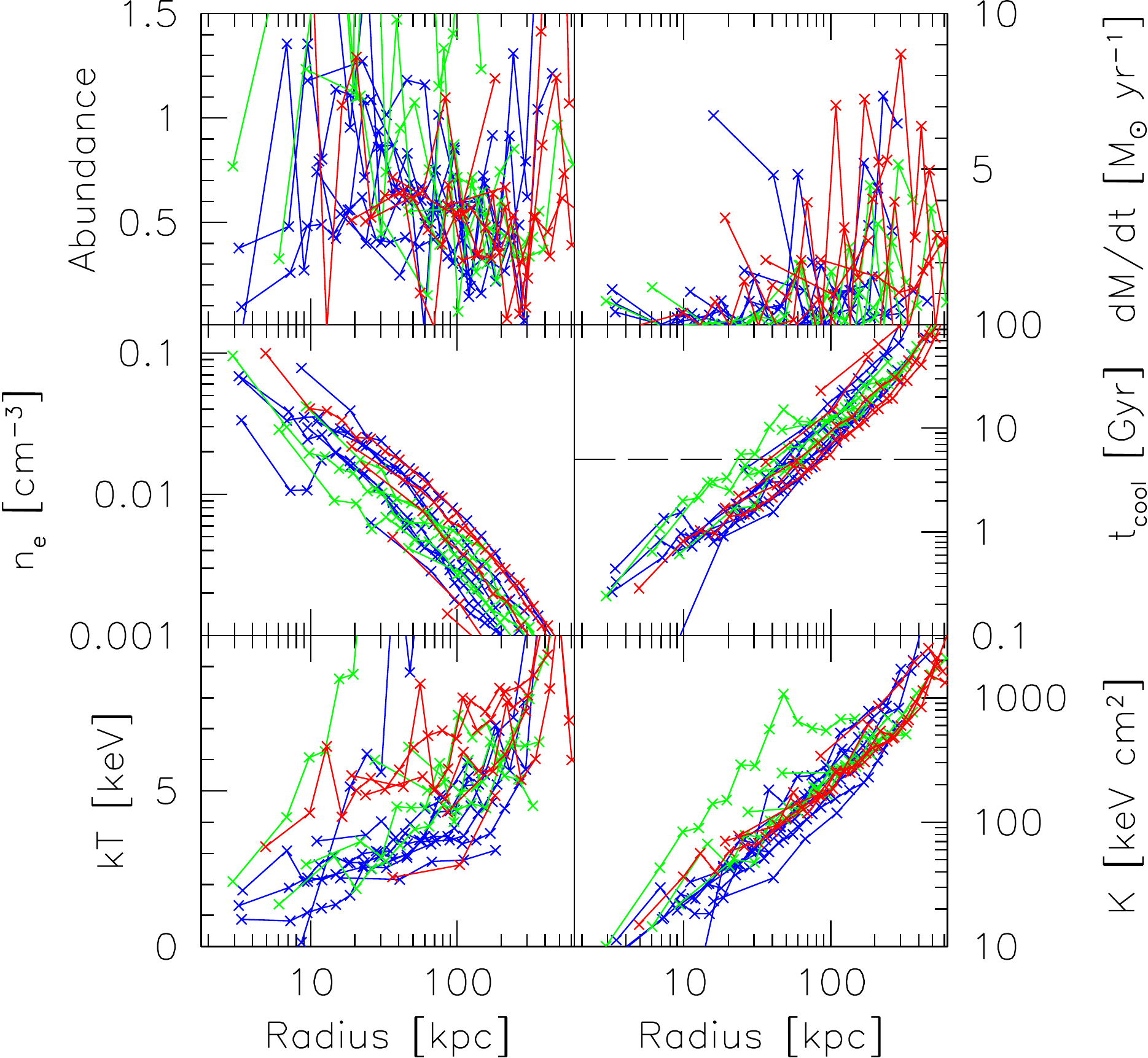}
\caption{X-ray derived radial profiles for abundance, electron density
(n$_e$), temperature (kT), specific entropy (K~=~kT$\cdot$n$_e^{-2/3}$
keV~cm$^2$), cooling time (t$_{cool}$ =
10$^8$$\cdot$(K/10~keV~cm$^2$)$^{3/2}$$\cdot$(kT/5~keV)$^{-1}$ Gyr) and mass
deposition rate (dM/dt). The dashed line in the right middle panel
represents a cooling time of 5 Gyr. The color scheme here is the same
as in Figure \ref{uvha}. The lower two panels show clearly that the
clusters with H$\alpha$ filaments (blue) have lower temperature and
entropy everywhere than those with nuclear or no H$\alpha$ emission
(green and red respectively).}
\label{profiles}
\end{center}
\end{figure}

Figure \ref{profiles} shows the radial profiles for various parameters
derived from the X-ray data: temperature (kT), density (n$_e$), metal
abundance, specific entropy (K~=~kT$\cdot$n$_e^{-2/3}$~keV~cm$^2$),
cooling time (t$_{cool}$~=~10$^8$$\cdot$(K/10 keV
cm$^2$)$^{3/2}$$\cdot$(kT/5 keV)$^{-1}$~Gyr) and mass deposition rate
(dM/dt). There appears to be very little difference in the shape or
absolute normalization of the density profile between clusters with
and without optical filaments. Likewise, while there is a large spread
in the values, there appears to be very little correlation between the
radial abundance profile and the presence of optical
emission. Interestingly, there appears to be a very strong correlation
between the absolute level of the temperature profile and our filament
type classification. That is, for clusters with extended, bright
optical filaments (type I), the temperature is lower at all radii. For
those with nuclear or no emission (types II or III), the temperature
is higher at all radii. Since the presence of H$\alpha$ filaments has
been shown to be enhanced in cooling flow clusters (Edwards et al
2007), one would expect the core temperature to be correlated with the
presence of filaments. We find that this correlation extends from the
core to much larger radius, such that the temperature at all radii
from the very center out to the cooling radius is lower in clusters
with extended H$\alpha$ filaments.

The motivation for producing these radial profiles was to look for a
correlation between the entropy profile and the presence of optical
filaments, which was alluded to by Donahue \etal (2006). However,
while previous authors (e.g. Cavagnolo \etal 2008) have found that the
central entropy, K$_0$, is much lower in clusters with H$\alpha$
emission, we find that the entropy is lower everywhere interior to the
cooling radius for clusters with extended H$\alpha$ emission. Naively,
one would expect strong AGN feedback and cluster-cluster mergers to
raise both the local temperature and entropy, thus impeding the
formation of H$\alpha$ filaments. However, simulations have found
that, under the correct circumstances, AGN feedback can lead to the
formation of cold filaments as radio-blown bubbles rise to large radii
(e.g. Fabian \etal 2003, Revaz et al. 2008, Pope \etal 2010).

\begin{figure}[p]
\centering
\includegraphics[width=0.9\textwidth]{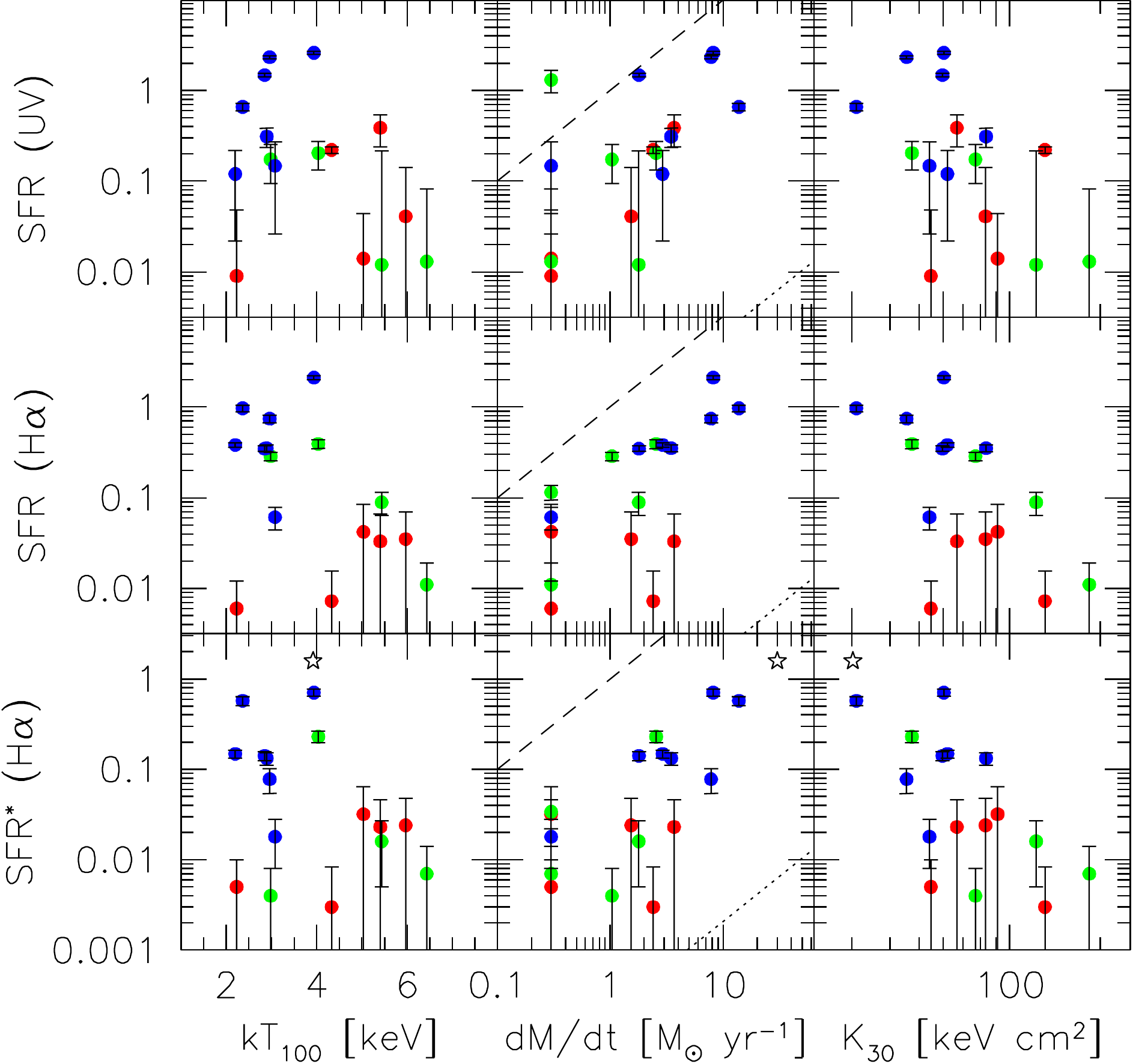}
\caption{UV and H$\alpha$ determined star formation rates, based on
Kennicutt (1998), against the average temperature in the inner 100 kpc
(kT$_{100}$), specific entropy at 30 kpc (K$_{30}$), and the
integrated mass deposition rate (dM/dt). The color scheme here is the
same as in Figure \ref{uvha}. The SFRs in the lower panels have had
the nuclear contribution removed (SFR$^*$). The open star in the lower
panels refers to Perseus A (Conselice \etal 2001, Sanders \etal
2004). The four points with dM/dt = 0.3 M$_{\odot}$~yr$^{-1}$ have no
measurable cooling flow and are plotted as upper limits. The diagonal
dashed line represents the limit where all of the cooling X-ray gas
turns into stars, while the dotted line is the case where all of the
cooling X-ray gas is made up of hydrogen which recombines only once
(Fabian \etal 1984). A strong correlation is seen between the
H$\alpha$ luminosity in filaments, and both the mass deposition rate
and the average cluster entropy.}
\label{uvha_xray}
\end{figure}  	  


We observe that both the temperature and entropy are lower in the
inner $\sim$ 100 kpc for clusters with observable H$\alpha$ filaments
(Figure \ref{profiles}). In order to quantitatively compare this
phenomenon to the presence of ionized filaments, we consider
quantities derived with this scale in mind.  Figure \ref{uvha_xray}
shows the comparison of the NUV and H$\alpha$ luminosities with the
average temperature inside of 100~kpc (kT$_{100}$), the specific
entropy at 30~kpc (K$_{30}$) and the mass deposition rate (dM/dt). The
UV and H$\alpha$ luminosities are shown in terms of their inferred
star formation rates, derived using Kennicutt (1998), so that the
comparison to dM/dt is straightforward. However, these values are
linearly proportional to the luminosity and so any observed
correlation is independent of this choice of scaling.

In Figure \ref{uvha_xray}, we see a strong correlation between the UV
or H$\alpha$ luminosity and both the entropy and dM/dt over three
orders of magnitude. Previous works (e.g. Cavagnolo \etal 2008) have
pointed out that systems with lower core entropy tend to have
UV/H$\alpha$ emission, however we show that the entropy outside of the
core is, potentially, an even better tracer of H$\alpha$ emission. We
prefer this diagnostic since several of our clusters have relatively
low surface brightness X-ray emission, making it difficult to
constrain the entropy in the core where the signal-to-noise is
low. The lower entropy systems tend to have brighter H$\alpha$
filaments, with Abell~1795, Sersic~159-03 and Perseus A occupying the
lowest-entropy and highest H$\alpha$ luminosity regime. The
correlation between dM/dt and the H$\alpha$ luminosity suggests a link
between the cooling flow and either the presence or ionization of the
warm gas.  The observed H$\alpha$ luminosity is $\sim$1-2 orders of
magnitude too high for it to be due to recombination from the hot
X-ray gas, assuming a single H$\alpha$ photon per recombination. On
the other hand, even if all of the H$\alpha$ is due to star formation,
there is not enough ongoing star formation to account for all of the
gas in the cooling flow. However, one should be cautious when using
the X-ray derived dM/dt since they do not represent the true mass of
gas cooling out of the X-ray, since it is still unclear how cooling
proceeds below soft X-ray energies (Peterson \& Fabian 2006). Note,
nevertheless, that there appears to be a temperature threshold of
$\sim$ 5 keV for ionized filaments, below which nearly all clusters in
our sample have extended H$\alpha$ emisison. It is unclear at present
whether there is a lower temperature threshold below $\sim$ 2 keV,
which is the domain of galaxy groups.

\begin{figure}[p]
\begin{center}
\includegraphics[width=0.9\textwidth]{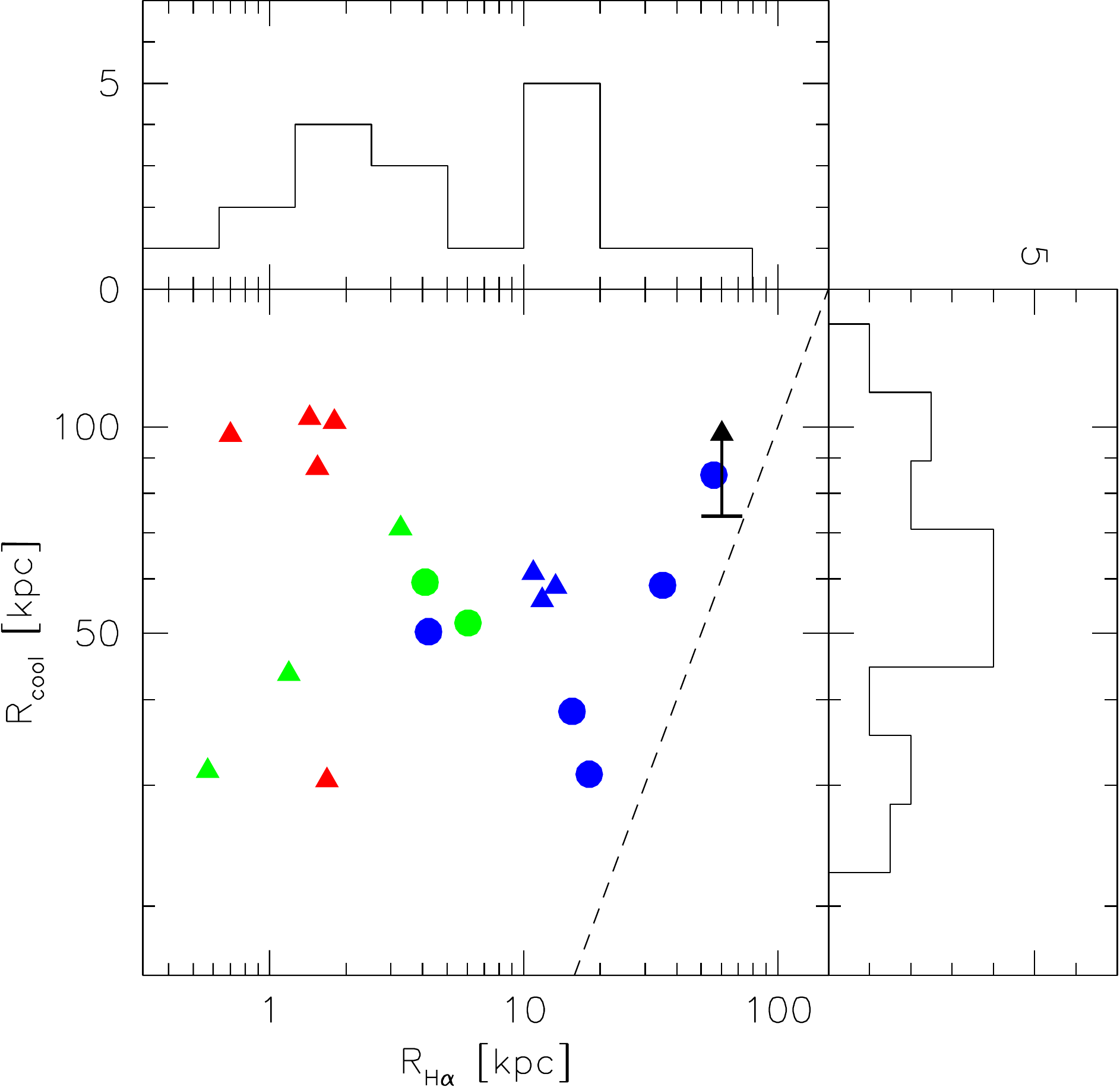}
\caption{Correlation and distributions of R$_{cool}$, the radius at
which the cooling time of the ICM reaches 5Gyr; and R$_{H\alpha}$, the
largest radius at which we detect H$\alpha$ emission. For the cases
with no detected H$\alpha$ emission an upper limit of the seeing FWHM
has been used. The dashed line refers to the one-to-one case, while
the point colors are consistent with Figure \ref{uvha}. The shapes of
the symbols refer to the NUV/H$\alpha$ ratio: circles have
NUV/H$\alpha$~$<$~10$^{-13.2}$~Hz$^{-1}$, while triangles have
NUV/H$\alpha$~$>$10$^{-13.2}$~Hz$^{-1}$. Note that, in the burst
formation scenario this corresponds to age, with circles representing
younger and triangles older stars. There appears to be an upper limit
on the radius of H$\alpha$ filaments, corresponding to the cooling
radius. The black point represents a lower limit estimate on the
cooling radius of Perseus A (Fabian \etal 2000, Conselice \etal
2001). The effect of decreasing the cooling time used to defined
R$_{cool}$ is to move points down in this plot.}
\label{radii}
\end{center}
\end{figure}

\begin{figure}[p]
\begin{center}
\includegraphics[width=0.9\textwidth]{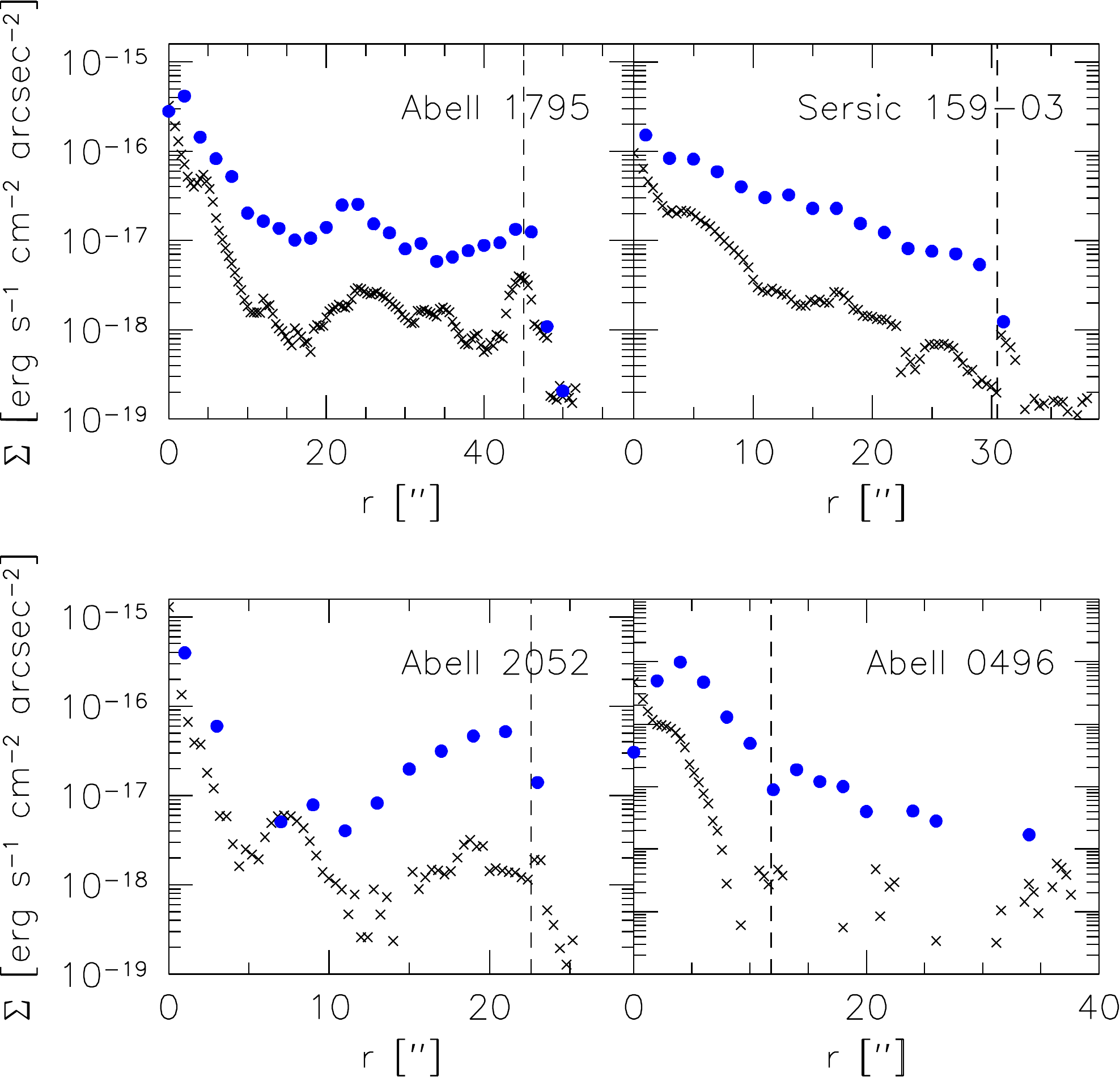}
\caption{H$\alpha$ surface brightness profiles for the three clusters
with R$_{H\alpha}$/R$_{cool}$ $\sim$ 1 and one cluster with
R$_{H\alpha}$ $<<$ R$_{cool}$ (Abell~0496). The black crosses
represent the azimuthally averaged profile, while the blue circles are
the surface brightness along a $\sim$ 5$^{\prime\prime}$ wide cut,
oriented along the most extended filament. The vertical dashed line
represents the location of R$_{H\alpha}$, as measured by eye. The
sharp drop-off of surface brightness at R$_{H\alpha}$ suggests that
this is truly the edge of the filament, independent of the depth of
our exposure.}
\label{sbprofs}
\end{center}
\end{figure}

From the radial X-ray profiles and the 2-dimensional H$\alpha$ data,
we are also able to derive two interesting scale radii. These are:
R$_{cool}$, the radius at which the cooling time of the X-ray gas is
equal to 5 Gyr, and R$_{H\alpha}$, the maximum extent of the H$\alpha$
filaments. The relationship between these radii is shown in Figure
\ref{radii}. While there is a large observed range in the radius of
the H$\alpha$ emission, there appears to be a hard edge at R$_{cool}$
-- we never observe H$\alpha$ emission beyond the radius where the
X-ray cooling time is $>$ 5 Gyr.  This is a strong indication that the
H$\alpha$ emission is intimately linked to the X-ray cooling flow. The
fact that the data appear to prefer a certain timescale (5 Gyr) for
the cooling radius is a new and puzzling discovery. Interestingly,
recent studies have shown that beyond z~$\sim$~0.5 the frequency of
clusters with cool cores is significantly lower than in the local
Universe (see e.g. Vikhlinin \etal 2007, Santos \etal 2008). This
redshift corresponds to a lookback time of 5~Gyr, which provides
further motivation for our choice of cooling radius. Thus, if cooling
flows were unable to begin until 5 Gyr ago as the evidence suggests,
we would expect the radius at which t$_{cool}$~=~5~Gyr to be a natural
boundary for the H$\alpha$ filaments.

Additionally, if we adopt the idea of stars forming in bursts, as
indicated by the location of the points in Figure \ref{uvha}, our data
show that the emission from the youngest stars tends to be much more
extended, while the emission from stars formed in a burst $\sim$ 10
Myr ago tend to be concentrated in a nuclear region, lending some
support to a radially-infalling scenario.

Since we are claiming a link between the cooling radius and the
maximum radius of H$\alpha$ emission, it is important to assess
whether we are truly seeing all of the H$\alpha$ emission, or
if a deeper exposure would yield R$_{H\alpha}$/R$_{cool}$ $>$ 1. In
Figure \ref{sbprofs}, the H$\alpha$ surface brightness profiles of the
three clusters with R$_{H\alpha}$/R$_{cool}$ $\sim$ 1 are shown. At
R$_{H\alpha}$, there is a clear drop-off in the surface brightness in
all three systems, suggesting that this is a real limit, not a
function of our exposure depth. For clusters with R$_{H\alpha}$ $<$
0.5R$_{cool}$ (e.g. Abell~0496, Figure \ref{sbprofs}, the edge of the
filament blends into the background much more smoothly, suggesting
that these filaments may in fact extend even further.
		  
\subsection{Properties of X-Ray Gas On and Off of Filament}
Using smoothed H$\alpha$ images as masks, we were able to extract
CXO spectra on and surrounding the warm ionized filaments, as
described in \S2.3.2. The off-filament spectra were fit with a
single-temperature model, while two cases were considered for the
on-filament spectra. In order to quantify the so-called ``iron bias''
(B\"{o}hringer \etal 2009), which results from fitting a
multi-temperature spectrum with a single-temperature model, we try
fitting both one-temperature and two-temperature models to the
on-filament spectra. A two-temperature model is clearly motivated if
we assume that the ICM conditions are different on-filament, since it
allows us to account for the ICM seen in projection in front of and
behind the filaments. The S/N of the on-filament spectra are typically
insufficient to constrain the temperatures, abundances and relative
normalization in a two-temperature model, so we opt to freeze the
temperature and abundance for one of the two models to the
off-filament conditions.

As we mentioned in \S  2.3, measuring the electron density on-filament
requires  an   assumption  about   the  geometry  of   the  extraction
region. For circular annuli, this is straightforward since we assume a
spherical cluster. Since  we can not be certain  about the geometry of
the filaments,  we consider two  limiting cases: 1) the  filaments are
thin  tubes and  extend no  further  than the  width of  our MMTF  PSF
($\sim$ 1 kpc) into the plane  of the sky; 2) the filaments are sheets
seen edge-on  with length equal to  the observed size  of the cluster.
For the first case we model the spectrum with a two-temperature plasma
to represent the thin filament  embedded in the surrounding ICM, while
the second  case is modeled as  a single-temperature slab  of gas. The
fact that we only see  filaments that are unresolved in width suggests
that  the   filaments  are  indeed   thin  and  not  sheets   seen  in
projection.  However  we  proceed   under  the  assumption  that  both
geometries  are  reasonable,   thus  bounding  the  ``true''  solution
somewhere in between.

\begin{figure}[p]
\centering
\includegraphics[width=0.9\textwidth]{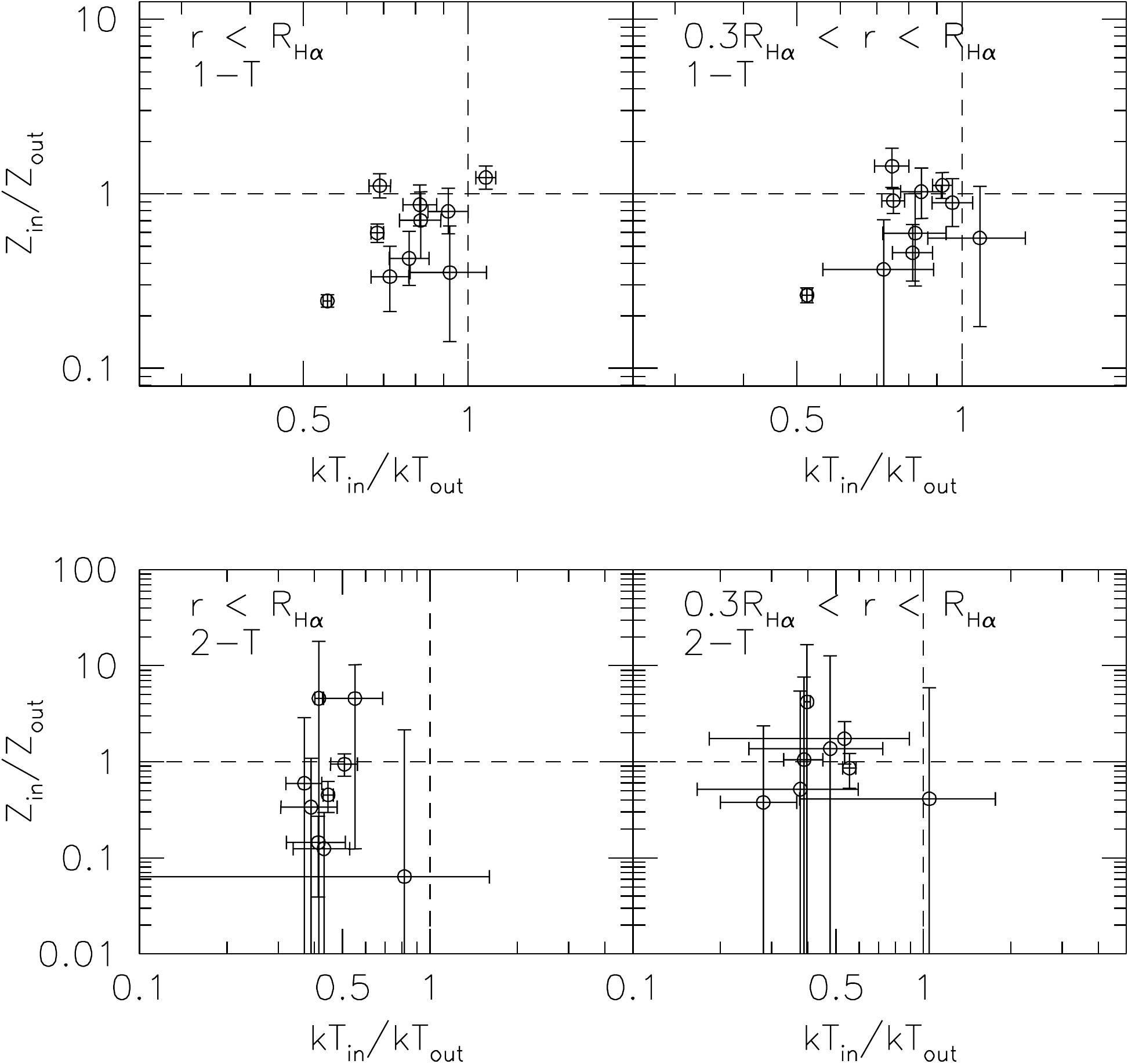}
\caption{Upper left panel: Ratio of metallicity and temperatures of
the X-ray gas coincident with H$\alpha$ filaments to that of the
surrounding ICM for a single-temperature model. The dashed lines
represent equal temperature and abundances inside and outside of the
filaments. The filaments tend to have systematically lower
temperatures and abundances in all cases with the exception of
Abell~0780. Upper right panel: Similar to the left panel, except that
only the outer 70\% of the filament in radius is considered (see
\S2.3.2). Lower panels: Similar to the upper panels, except using a
2-temperature model to fit the on-filament spectrum. Note that this
results in significantly lower on-filament temperatures and overall
larger errors in both fitting parameters.}
\label{zinout}
\end{figure}

Fortunately, the assumption of geometry does not affect the
calculation of the abundance or temperature. The on-versus-off
filament temperature and abundance ratios are shown in Figure
\ref{zinout}, for both single-temperature and two-temperature
models. Immediately obvious from this figure is the fact that the
filaments tend to be cooler than the immediate surroundings -- the
typical temperature in the filament is $\sim$ 50\% of that in the
nearby gas. The fact that the X-ray gas near the H$\alpha$ filaments
is significantly cooler than the surrounding ICM is further evidence
that the H$\alpha$ filaments may be taking part in the cooling
flow. Additionally, the on-filament metallicity tends to be
significantly lower, reaching values less than $\sim$ 40\% of the
off-filament values for about half of our systems. This does not
appear to be the case for the 2-temperature model applied to the
``outer filament'' spectrum, suggesting that the low on-filament
metallicity may be a result of the iron bias (B\"{o}hringer \etal
2009). However, this is also the lowest S/N of the four scenarios
considered. Thus, deeper data are required in order to say for certain
what the true abundance gradient between the on- and off-filament
spectra is. There are very few physical processes that decrease
metallicity so, if the on-filament abundance is in fact lower with
respect to the surrounding ICM, it is more likely that the
low-metallicity gas has been transported from elsewhere.

\begin{figure}[p]
\centering
\includegraphics[width=0.8\textwidth]{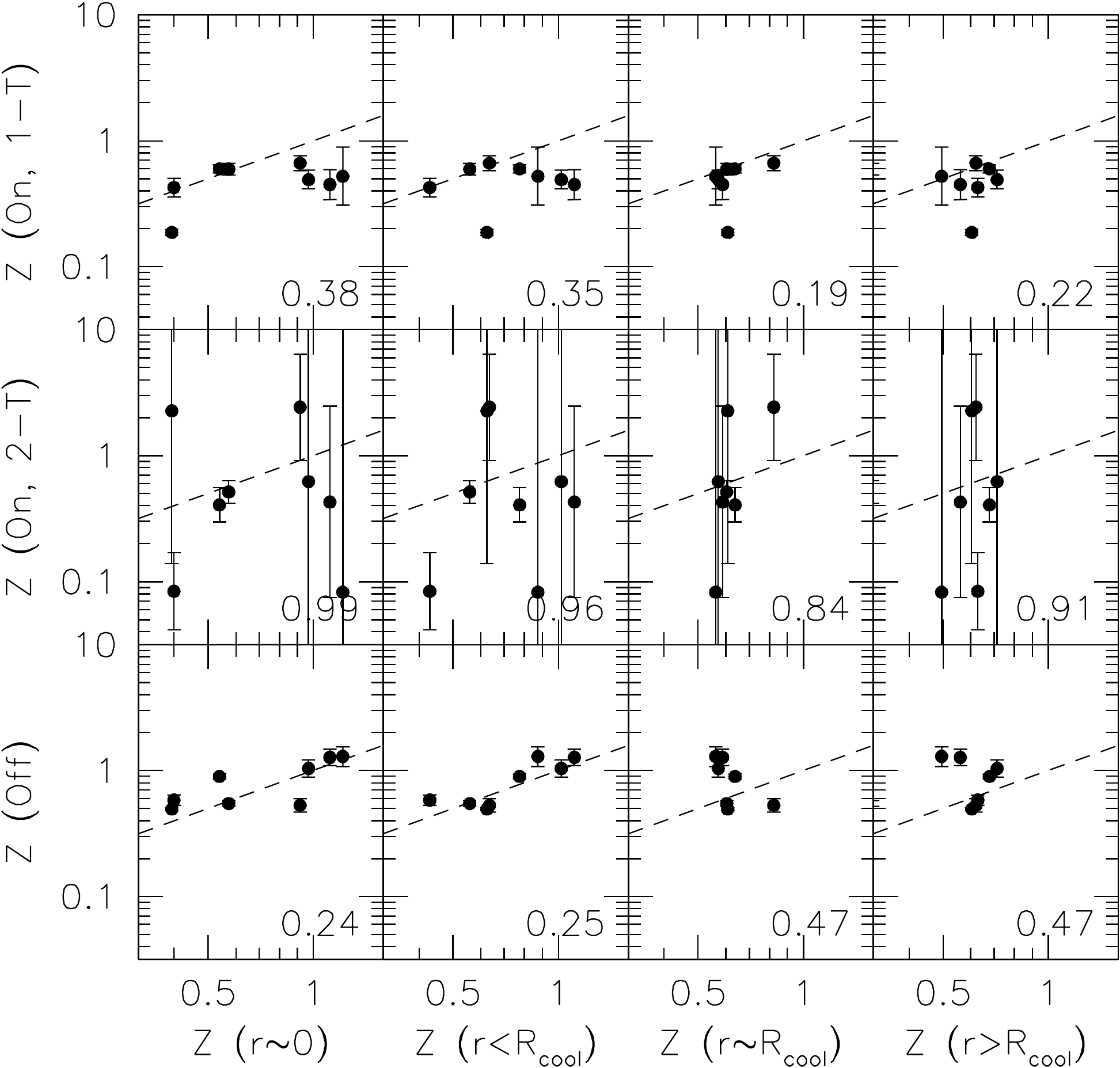}
\caption{Upper panels: Metallicity of X-ray gas coincident with the
H$\alpha$ filaments as a function of the metallicity in various radial
bins. The number in each panel is the standard deviation around the
line of equality. The metallicity of the in-filament gas matches more
closely the metallicity of the ICM at the cooling radius than in any
other radial region. The lowest point in the upper panels, which is an
outlier in all four panels, is Abell~2052. For a discussion of why
this cluster is an outlier, see the section on indixidual clusters in
the Appendix. Middle panels: Same as upper panels, but now considering
a two-temperature fit to the on-filament spectrum. The additional free
parameters yields a poorly constrained in-filament abundance. Again,
the abundance at the cooling radius appears to match best the
in-filament abundance, but the difference is now only marginal. Lower
panels: Same as above, but for the X-ray gas outside of the H$\alpha$
filaments.}
\label{abundances}
\end{figure}

\begin{figure}[p]
\centering
\includegraphics[width=0.9\textwidth]{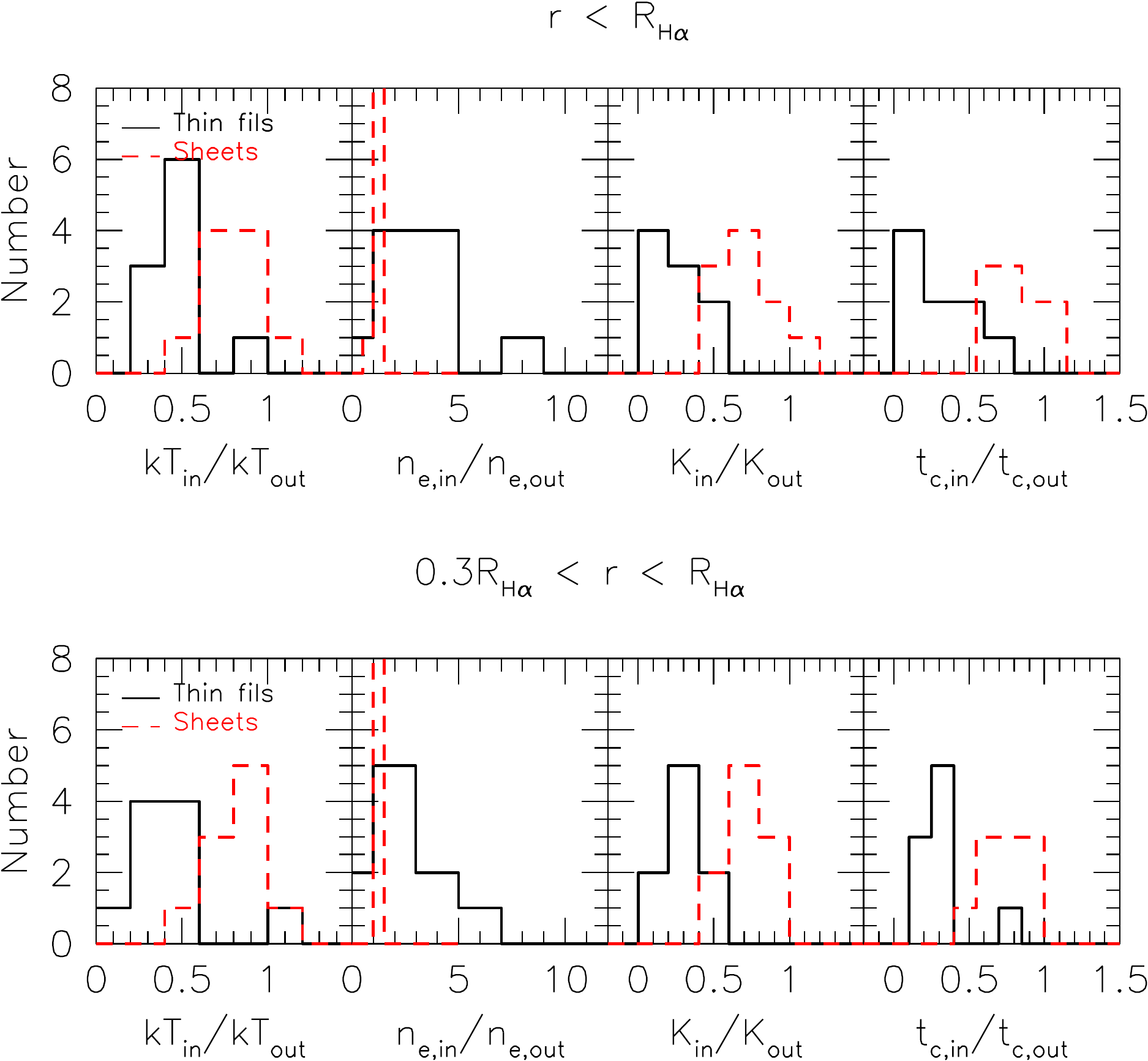}
\caption{Upper panels: Distribution of temperature, electron density,
entropy and cooling time ratios in and out of the H$\alpha$
filaments. The two line types bracket the extreme cases of the
filament geometry: solid black lines are the thin filament case, which
is modeled with a two-temperature plasma, while the red dashed line are
the case of single-temperature sheets seen edge-on. The in-filament
gas shares similar properties with the cooling flow, namely that the
X-ray gas has a cooling time of $\sim$20\% that of the surrounding
ICM. Lower panels: Similar to the above panels, but now considering
only the outer 70\% of the filaments in radius. }
\label{fils}
\end{figure}


A hint as to the origin of this low-metallicity gas can be found if we
look at the 5 clusters with the largest on-off filament metal
abundance contrast. With the exception of Abell~2052, all of these
clusters have very strong radial abundance gradients. On the other
hand, those clusters which do not have a strong on-off filament metal
abundance contrast have relatively flat radial abundance gradients. To
further drive this home, we show in Figure \ref{abundances} the
relationship between the on- and off-filament abundance and the
abundance in four regimes: (1) r $\sim$ 0, (2) r $<$ r$_{cool}$, (3) r
$\sim$ r$_{cool}$ and, (4) r $>$ r$_{cool}$. This figure shows that,
in general, the metallicity in the filaments matches best with the
metallicity at r$_{cool}$. For those four clusters with strong on-off
filament abundance gradients (excluding Abell~2052), the on-filament
abundance matches almost exactly the local abundance at
r$_{cool}$. For the remaining four clusters, the on-filament abundance
looks roughly the same as the abundance at all radii, due to the
relatively flat abundance gradient. If we consider a two-temperature
model for the on-filament spectrum, the correlation between the
on-filament metallicity and the metallicity at the cooling radius
becomes considerably less significant. Again, more data is needed in
order to achieve a high enough on-filament S/N in order to constrain
the abundance in a two-temperature model. A direct link between the
on-filament metallicity and the metallicity outside of the cooling
radius would further strengthen the connection between the H$\alpha$
filaments and the cooling flow. Furthermore, it would offer some
evidence against the common hypothesis that the gas in the H$\alpha$
filaments comes from the central region, either from radio-blown
bubbles (i.e. Revaz \etal 2008) or from gas sloshing (i.e. Johnson
\etal 2010). While in some cases the match between the core and and
on-filament abundances is good, it is overall worse than the match to
the abundance at R$_{cool}$. However, there are two biases which are
likely conspiring to increase the uncertainty in the core abundance
measurement. First, we expect the core abundance to be most affected
by the iron bias, implying that the core abundances which we quote are
likely lower limits. At the same time, our coarse binning in the
central region (chose to ensure a high S/N deprojection) means that
any strong dip in the central abundance will be smeared out, leading
to a slightly higher measured abundance. As an example, for Abell~4059
we find a dip in the abundance profile in the center, with a central
value of $\sim$ 1.0, while Reynolds \etal (2008) quote a value of
$\sim$ 0.3 due to their higher resolution. Using a 2-component model,
Reynolds \etal find that the core abundance dip changes to an excess,
with a central value of $\sim$ 2.5. Thus, it is clear that the value
of the central abundance is highly dependent on the model and binning
chosen, and is therefore uncertain.

In addition to the X-ray temperature and metallicity in the filaments,
we can also measure various additional properties such as electron
density, entropy and cooling time, as shown in Figure \ref{fils}.  The
distribution of these properties offers further evidence for a link
between the X-ray cooling flow and the H$\alpha$ emission. Assuming
the thin-filament geometry, the density inside of the filament is
typically a factor of a few higher than the surrounding ICM. Coupled
with the temperature decrease mentioned above, this yields drastically
lowered entropy and cooling times inside of the filament. With cooling
times ranging from 10\%-40\% of the surrounding ICM, these regions
are, by definition, part of the cooling flow. It should also be noted
that, even if we assume that the geometry is that of
single-temperature sheets seen in projection, the cooling time and the
entropy are lower overall inside of the filaments.

The overall results from extracting X-ray spectra coincident and
anti-coincident with the observed H$\alpha$ filaments tell an
interesting story. 
The H$\alpha$ filaments appear to reside in a portion of the ICM
with higher densities that is cooling more rapidly than its
surroundings, resulting in lower temperatures. The metallicity inside
of the filaments resembles more closely the metallicity at the cooling
radius than the adjacent off-filament gas, albeit with a considerable
amount of uncertainty. These two results, coupled with the fact that
the observed H$\alpha$ emission seems to know about the cooling radius
(R$_{H\alpha}$ $\lesssim$ R$_{cool}$), suggests that the observed
H$\alpha$ filaments are coupled to the cooling flow.

\begin{figure}[p]
\centering \includegraphics[width=0.9\textwidth]{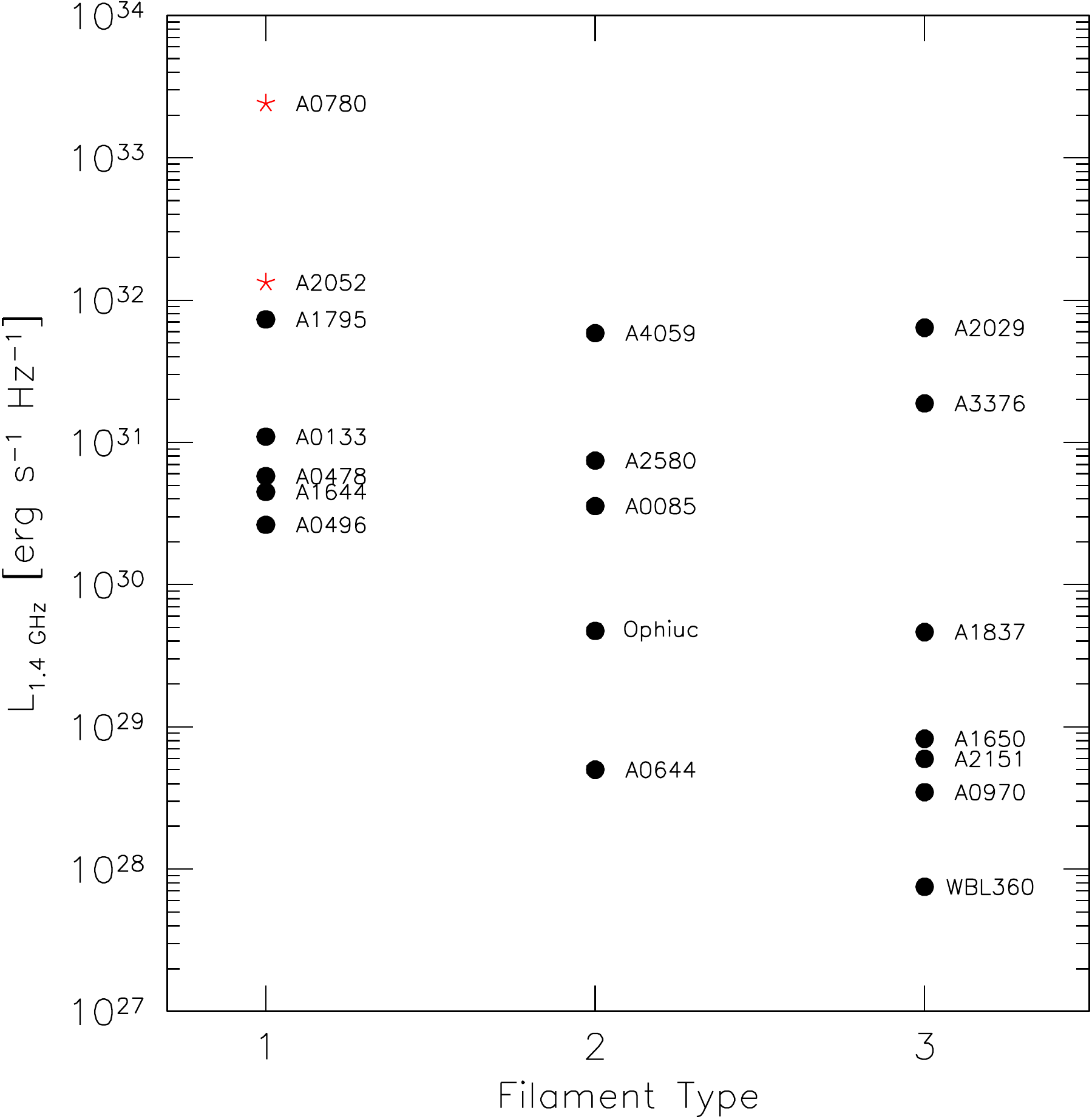}
\caption{1.4 GHz radio luminosity for all clusters detected in the NVSS, as a function of the filament type described in \S3.1. The red star-shaped points represent the two clusters with an X-ray point source detected in the center, while the filled black circles represent those clusters with no detected X-ray point source. There appears to be little correlation with the radio power and the presence of optical filaments, as 2 of the 6 brightest clusters in radio have no detectable H$\alpha$ emission whatsoever.}
\label{agn}
\end{figure}

\begin{figure}[htb]
\centering
\begin{tabular}{c c}
\includegraphics[width=0.7\textwidth]{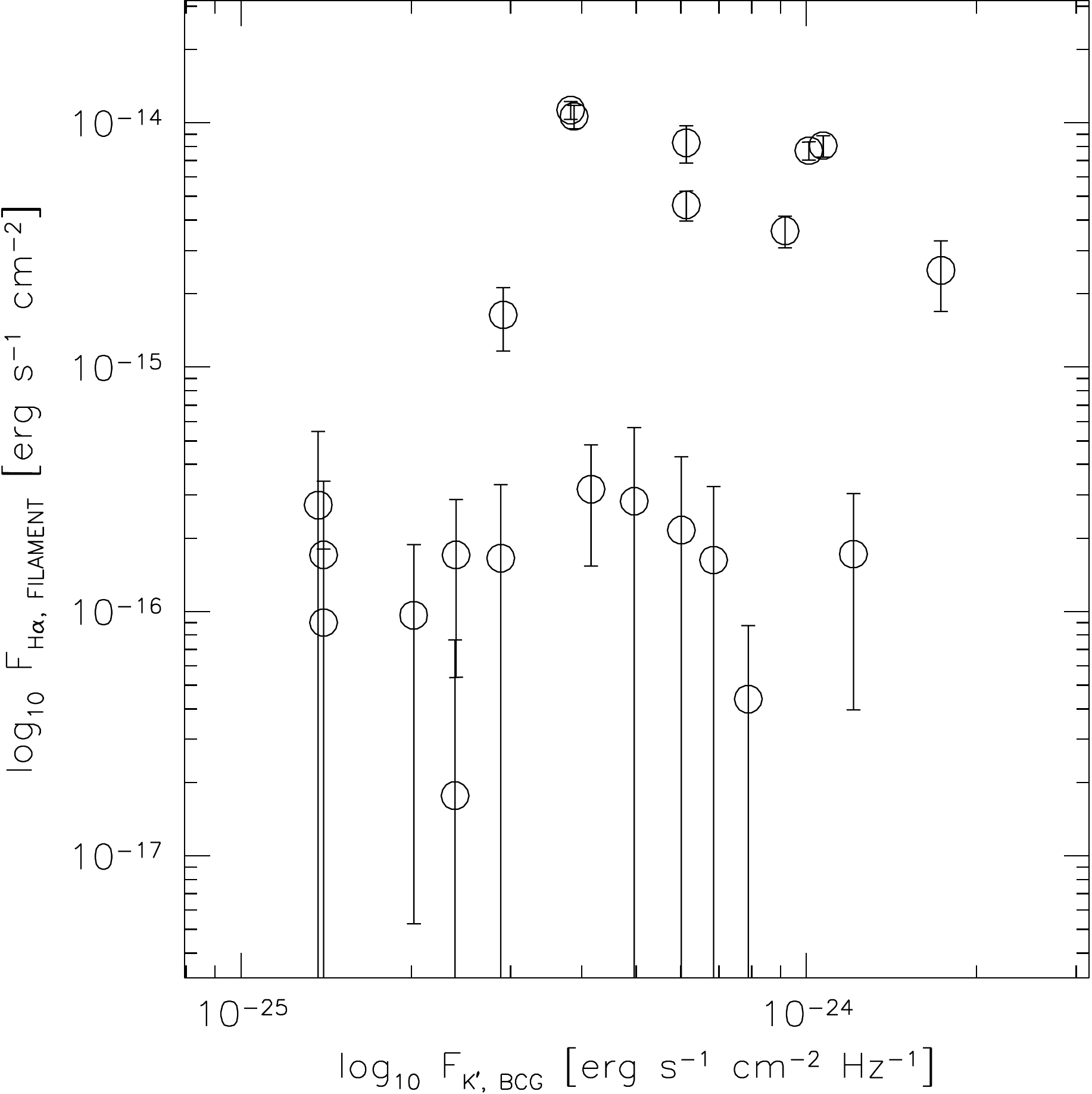}
\end{tabular}
\caption{H$\alpha$ flux contained in filaments (nucleus removed) versus the total BCG K$^{\prime}$-band flux from 2MASS. There is no obvious correlation between the brightness of the BCG and the amount of warm ionized hydrogen in filaments.}
\label{HaK}
\end{figure}

\subsection{Properties of BCGs}
As a method of probing whether or not these clusters harbor an AGN in
their central galaxies, we consider the 1.4 GHz radio power as well as
the presence of an X-ray point source in the cluster core. The
distribution of radio luminosities for our sample is shown in Figure
\ref{agn}. We note that all clusters with filaments have a
significant radio detection, while only 5/15 of the clusters without
filaments have a significant radio detection. Thus, the presence of
radio emission is not predictive of whether there will be extended
H$\alpha$ filaments. The two clusters with detected X-ray point
sources in the BCG, Abell~0780 (Hydra A) and Abell~2052 have large
cavities in their X-ray halos, suggesting that the AGN is strongly
influencing the surrounding medium. The H$\alpha$ in these two
clusters is very likely linked to this activity. If we consider the
remaining clusters, the correlation between radio power and the
presence of H$\alpha$ filaments becomes much weaker, suggesting that
the two phenomena are not directly tied to one another. Instead,
whatever weak correlation that is seen could be due to the fact that
cool core clusters, which tend to have H$\alpha$ filaments (Figure
\ref{uvha_xray}), will also have a more ready supply of fuel for the
central AGN, leading to increased radio feedback. Further discussion
of the implication of these, and other, results follow in the next
section.

We can also consider the relationship between the BCG itself and the
presence of optical filaments. Figure \ref{HaK} shows the total
K$^{\prime}$-band flux for the BCG versus the total H$\alpha$ flux
contained in filaments. The K$^{\prime}$-band flux is a good
approximation to the total stellar mass, with only a very small
deviation with galaxy color (Bell \& de Jong 2001). There is no
obvious correlation between the the galaxy brightness (or stellar
mass, approximately) and the presence of H$\alpha$ filaments,
suggesting that the observed filaments are related instead to the
X-ray properties of the cluster core.


\section{Discussion: Origin of the H$\alpha$ Filaments}
\begin{figure}[htb]
\centering
\includegraphics[width=0.9\textwidth]{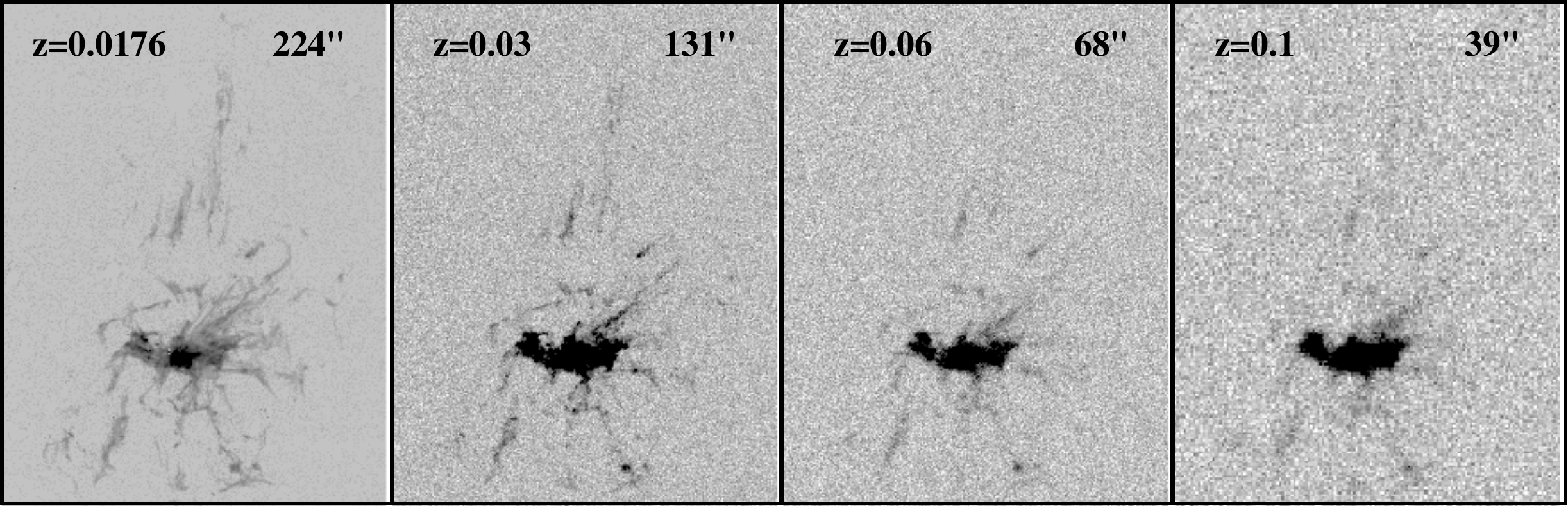}
\caption{Image of Perseus A from Conselice \etal (2001). The image quality has been degraded to simulate the effect of observing this cluster at varying distances. The redshift of observation is in the upper left corner, while the apparent size of the image in the horizontal direction is in the upper right. By the mean redshift of our sample, z $\sim$  0.06, most of the spectacular filaments have been lost, leaving only a few unresolved structures.}
\label{perA}
\end{figure}

In the previous sections we have provided several new clues to the
origin of the observed H$\alpha$ filaments:
\begin{itemize}
\item In clusters with filamentary H$\alpha$ emission, we tend to see
structure in the cooler (0.5--2.0~keV) X-ray gas. If there is no
structure in the X-ray gas, there is typically no accompanying
H$\alpha$ emission (Figures \ref{bigfig}, \ref{xray-rgb}).
\item There is a strong correlation between the H$\alpha$ flux
in filaments and both the X-ray determined cooling flow rate,
dM/dt, and cluster entropy over 3 orders of magnitude (Figure
\ref{uvha_xray}).
\item The extent of the H$\alpha$ filaments never exceeds the cooling
radius. This appears to be a hard limit to the radius of these
filaments (Figure \ref{radii}).
\item The metallicity of the X-ray gas coincident with the H$\alpha$
filaments is often lower than the non-coincident ICM at
the same radii. The on-filament X-ray metallicity is consistent with
that measured near the cooling radius or beyond (Figure \ref{abundances}).
\item The temperature and cooling time of the X-ray gas coincident with the H$\alpha$
filaments is significantly lower than that of the surrounding
ICM (Figure \ref{fils}).
\end{itemize}
These results suggest a direct link between the hot ICM and the warm
ionized filaments. However, before we discuss their implications, we
would be remiss if we did not mention the spectacular H$\alpha$
filaments detected in the nearby cooling flow cluster, Perseus. A
great deal of literature has been devoted to the study of these
filaments, which are seen in molecular gas (Salom\'{e} \etal 2008),
warm ionized hydrogen (Conselice \etal 2001) and X-ray (Fabian \etal
2006). However, while the observed filaments are indeed impressive, we
would argue that they are ``typical'' among what we classify as our
type I clusters. In Figure \ref{perA} we show the effect of placing
Perseus A at larger redshift, and simulating a 20-minute exposure
using the MMTF under typical observing conditions. By a redshift of
0.06, which is roughly the mean redshift of our sample, the majority
of the fine structure has been washed out, leaving only a small number
of detectable filaments. These redshifted images look similar in
morphology to our detections of filaments in the 7 clusters which we
label as type I. Additionally, we show in Figure\ \ref{uvha_xray} that
Perseus A occupies a similar space to the other clusters with
filamentary H$\alpha$ in temperature, cooling rate and
temperature. The total luminosity in H$\alpha$ is significantly higher
than in all other clusters, however this is probably again partially
due to its proximity: if a filament completely disappears into the
background we will choose a smaller aperture and, thus, measure a
smaller total luminosity.

We finish by discussing the results summarized above in the context of
several current theories which attempt to explain the presence of the
observed H$\alpha$ filaments: buoyant radio bubbles, gas sloshing in
the central potential, accretion of gas-rich galaxies, filamentary
cooling flows, ICM conduction and magnetic fields.

\subsection{Origin of the Cool Gas}

\subsubsection{Buoyant Radio Bubbles}
In an effort to explain how an AGN can deposit energy into the ICM and
quench cooling, much work has been focused on the evolution of bubbles
blown in the ICM by radio jets (e.g. Reynolds \etal 2005,
Vernaleo\etal 2007, Revaz \etal 2008). Due to the local density
contrast, these bubbles are buoyant and will rise to larger radius,
transporting energy from small radii to large. While rising, the shape
of the bubble evolves and can leave behind a trail of cooler gas that
morphologically matches the observed H$\alpha$ filaments in Perseus A
(Reynolds \etal 2005, Revaz \etal 2008). Radio bubbles have been observed
in several clusters to date, including some in our sample
(e.g. Abell~0780, Abell~2052), which lends further support to this scenario.

This scenario is consistent with the low temperatures seen in the
X-ray ICM coincident with the H$\alpha$ filaments and with the fact
that all of the clusters with extended filaments have non-zero 1.4 GHz
flux. Some clusters (e.g. Abell~0780, Abell~2052) show H$\alpha$
emission along the edges of known radio-blown bubbles lending further
support to this scenario. It is unclear exactly how the strong
correlation between the H$\alpha$ radius and the cooling radius fits
in this scenario, although it could be that this relation is tied to
the ionization and not to the source of the gas. The low ICM abundance
coincident with the H$\alpha$ filaments appears to match well with the
abundance beyond the cooling radius, but it may also match the core
abundance, which previous works (e.g., Sanders \etal 2002, 2007) have
found to be metal-poor in several nearby clusters.

\subsubsection{Gas Sloshing}
The gravitational disruption of a cluster core by mergers or
interactions with other clusters or galaxies can lead to the ICM
``sloshing'' about the central potential, creating cold fronts that
have been observed in several clusters to date (e.g. Zuhone \etal
2009, Johnson \etal 2010). While this is certainly an important
mechanism for heat transport and is very likely ongoing in several
clusters with observed cooling fronts, it does not seem to be a
plausible general explanation for the presence of warm filaments. The
morphology of the H$\alpha$ filaments in our sample tends to be
radial, with very few exceptions. The gas sloshing scenario predicts
cool, low-entropy gas along fronts which are aligned perpendicular to
the radial direction. The process of sloshing in the cluster core
should also smooth out any abundance gradients, however we detect
possible abundance contrasts between the X-ray gas on-filament and
off-filament. Finally, the magnitude of the gas sloshing should be
dependent on the strength of the merger or disruption event. However,
we see a clear limit to the length of the filaments (R$_{H\alpha}$ $<$
R$_{cool}$), which is inconsistent with the sloshing picture.

\subsubsection{Accretion of Younger Galaxies/Groups}
While the direction of some of the observed filaments seemingly
coincides with the location of smaller galaxies (e.g.\ Abell~0478,
Abell~0496, Abell~0780), it is unlikely that the warm gas has been
stripped from these galaxies. If this were the case, one would expect
the filaments to extend all the way to the satellite galaxy, in some
cases, which is not seen. Likewise, we would not expect a single
galaxy, or small group of galaxies, to have a strong effect on the
distribution of the hot X-ray gas.  However, much of the structure
seen in H$\alpha$ is seen also in the Chandra X-ray
images. Additionally, the strong correlation between the mass
deposition rate and cluster entropy would be unexplained by such a
formation method.

A back-of-the-envelope calculation seems to rule out this mechanism as
the source of the observed H$\alpha$ filaments. The typical H$\alpha$
luminosity that we observe in filaments is $\sim$ 5$\times$10$^{40}$
erg s$^{-1}$, which corresponds to $\sim$ 3.7$\times$10$^{52}$
recombinations per second (assuming case B recombination). If we
assume that the diffuse ionized medium in the stripped galaxy is pure
hydrogen and is being stripped from a galaxy like our Milky Way, and
then recombining only once as it cools in the absence of ionizing
radiation, this corresponds to $\sim$ 950 M$_{\odot}$ yr$^{-1}$ of
recombining gas. Assuming a total diffuse HII mass of 10$^9$
M$_{\odot}$ for a typical massive spiral (Lequeux 2005), this implies
that all of the stripped gas would recombine in $\sim$ 1 Myr. Thus, in
order to consistently produce warm ionized filaments via gas
stripping, the BCG would have to strip the gas disk from a spiral
galaxy every few Myr, which is unreasonably short, given the typical
cluster crossing times of $\sim$ 1 Gyr. Alternatively, if we invoke an
additional ionization source once the gas has been stripped from the
galaxy (e.g. X-ray conduction from the ICM), a smaller amount of
stripped gas could lead to the observed fluxes of H$\alpha$. We will
revisit this issue in much greater detail in subsequent papers.

\subsubsection{Filamentary Cooling Flows}
In this scenario, the observed H$\alpha$ filaments directly trace the
cooling flow (i.e. Cowie \etal 1980; Fabian \etal 1984; Heckman \etal
1989). The X-ray ICM begins to cool dramatically at the cooling
radius, while falling inwards at roughly the free-fall speed. As the
gas cools, small density contrasts in the ICM grow and thin,
high-density filaments form. Once sufficiently cool, the gas in these
filaments can be re-ionized via star formation, collisional heating
(e.g. Ferland \etal 2009) or drag heating (e.g. Pope \etal 2008). This
scenario is consistent with H$\alpha$ filaments extending to
R$_{cool}$, with the observed correlation between X-ray and H$\alpha$
properties and with the low metallicity and high cooling rates found
in the X-ray gas coincident with H$\alpha$ filaments. Additionally,
the fact that we only see H$\alpha$ filaments out to a radius where
gas can cool in 5 Gyr agrees well with observations that the frequency
of cool cores beyond a lookback time of 5 Gyr is significantly lower
than the local Universe. Thus, the cooling flow scenario is our
preferred formation scenario for the optical filaments.

The thinness of the observed H$\alpha$ filaments relative to their
extent (Table~A.1) is reminiscent of the thin, filamentary structure
seen in hydrodynamic simulations of cool gas flows (e.g.\ Hattori
\etal 1995, Pope \etal 2008, Ceverino \etal 2010). Figure 3 from
Ceverino \etal (2010) shows long ($>$ 50 kpc), thin ($<$ 10 kpc)
filaments of cold gas which resemble our observations of H$\alpha$
filaments. While the physical scales are slightly different, they
describe the process which cool ($<$~5$\times$10$^4$~K) filaments
embedded in a spherical halo of hot ($>$~3$\times$10$^5$~K) gas may
form. The formation of these cool filaments is used by the authors to
explain how star formation proceeds in spiral galaxies at high
redshift, but the physical processes should be similar for gas cooling
from the hot phase to the cool phase, regardless of whether we are
considering gas streaming onto an isolated spiral galaxy or a central
cluster galaxy.

In MV09 we argued that the most plausible scenario for the creation of
the long, thin filament in Abell~1795 was a cooling wake caused by a
cooling flow occurring around the moving cD galaxy. This claim was
made based on the observation by Oegerle \& Hill (2001) that
Abell~1795 has a significant peculiar velocity relative to the
cluster. However, we find that the second most asymmetric case,
Abell~0496, which also has filaments extending only in one direction, has
a negligible peculiar velocity. Likewise, several BCGs with no
H$\alpha$ emission (e.g., Abell~2029, Abell~2151) or nuclear-only
H$\alpha$ emission (e.g. Abell~2580) have a peculiar velocity as high,
or higher, than Abell~1795. Thus, while this may play a role in
determining the direction of the filaments, it is not likely the most
important factor in their formation.

This scenario makes the prediction that the filament material should
be in near free-fall. The jury is still out on this issue, as several
works (i.e. Hatch \etal 2007, Oonk \etal 2010) argue against radial
infall in H$\alpha$ filaments, while others (i.e. Lim \etal 2008,
Wilman \etal 2009) argue for it. Thus, a homogenous sample of
kinematic measurements is needed to complement this census of the warm
and hot medium and provide a strong argument either for or against the
radially-infalling cooling flow.

\subsection{Evidence for Conduction}

Conduction from the X-ray ICM is a promising scenario which could provide
the energy necessary to ionize the cooling gas. Nipoti \& Binney (2004)
predict that, if conduction is responsible for ionizing the H$\alpha$
filaments, there should be an excess of soft X-rays surrounding the
filaments. This is consistent with our findings that the soft X-ray
morphology matches well with the H$\alpha$ morphology and that the
temperature of the X-ray gas coincident with the warm filaments is
significantly cooler than the surrounding ICM. In an attempt to
address the plausibility of this scenario for ionizing the observed
H$\alpha$ filaments in our sample, we turn to a study of thermal
conduction in 16 cooling-flow clusters by Voigt \& Fabian (2004). Of
these 16 clusters, there are 5 which overlap with our sample:
Abell~0478, Abell~0780, Abell~1795, Abell~2029 and Sersic~159-03. Of
the five clusters in this overlapping sample, four show very extended,
filamentary H$\alpha$ emission. The conductivity inside of the cooling
radius for these four clusters ranges from
$\kappa_{eff}/\kappa_S$~=~0.3--2.3, while for the lone overlapping
cluster with no H$\alpha$ emission, the effective conductivity is
$\kappa_{eff}/\kappa_S$~=~0.2 (Voigt \& Fabian, 2004). Estimates of
the minimal value required for conduction to to transport heat from
the hot ICM to cooler regions varies from 0.01 (Churazov \etal 2001)
to 0.3 (Narayan \& Medvedev 2001).

The fact that we observe cooler X-ray gas coincident with the
H$\alpha$ filaments, coupled with the relatively large conductivity
measurements in clusters with filaments (albeit with a small sample
size) suggests that conductivity may be responsible for much of the
energy required to ionize these filaments. However, future work is
needed to expand on the number of clusters with effective conduction
measurements and to determine if the energetics of the warm filaments
match what we would expect from ICM conduction.

\subsection{Role of Magnetic Fields}
Given the typically large axial ratio of the H$\alpha$
filaments (length/thickness $\sim$ 30, according to Table A.1), magnetic
fields likely play a role in the formation of these filaments, in one
or more of several ways: (i) by channeling the inflow of clumps of gas
along field lines, giving them coherent structure (Fabian \etal 2003),
(ii) preventing the hot, turbulent ICM from shredding the filaments
(Hatch \etal 2007), (iii) to help the growth of thermal instabilities,
leading to thin high-density filaments (Hattori \etal 1995), and (iv)
suppressing conduction of heat from the ICM to the cooling gas (Voigt
\& Fabian 2004).

High-resolution X-ray spectra of these clusters could constrain the
level of turbulence in the ICM (e.g., Sanders \etal 2010). These
measurements, combined with the observed thinness of the H$\alpha$
filaments and MHD models could constrain the strength of the
intracluster magnetic field required to sustain such filaments, in a
similar manner to Loewenstein \& Fabian (1990). This would be a nice
consistency check on the magnetic field strengths derived from
rotation measures.


\section{Summary and Future Prospects}

This study represents the highest resolution survey of warm ionized
filaments in cooling flow clusters beyond 100 Mpc to date. Complex,
filamentary morphologies are seen in 8/23 (35\%) of our clusters, with
slightly extended or nuclear emission seen in additional 7/23
(30\%). We find a weak correlation between the total H$\alpha$
luminosity and the near-UV luminosity from young stars, which suggests
that photoionization from young stars may play a role in producing the
observed H$\alpha$. UV imaging of higher spatial resolution is needed
to spatially correlate the H$\alpha$ and UV emission, and to remove
the contribution to the UV luminosity from the AGN and/or central
starburst. While the H$\alpha$ and X-ray data support the buoyant
radio bubble hypothesis as a source of cool gas in some cases, we
believe that the cooling flow model does a better job of explaining
the properties of the ensemble of clusters. We observe correlations
between the soft X-ray morphology and the H$\alpha$ morphology, the
X-ray cooling rate and the H$\alpha$ luminosity contained in
filaments. We find evidence for a correlation between the X-ray metal
abundance of gas coincident with the H$\alpha$ filaments and at the
cooling radius, however more data are required in order to obtain
reliable on-filament abundance measurements. The maximum radial extent
of the H$\alpha$ emission and the cooling radius appear to be linked,
such that R$_{H\alpha}$ $<$ R$_{cool}$, where R$_{cool}$ is the radius
at which the X-ray gas can cool in less than 5~Gyr. This radius
appears to be physically motivated, as the choice of timescale agrees
well with the lookback time at which cooling flow clusters begin to
emerge. Additionally, we find that the X-ray gas coincident with
observed H$\alpha$ filaments is cooling on a much shorter ($\sim$
20\%) timescale than the surrounding ICM. These results suggest that
the H$\alpha$ filaments trace the X-ray cooling flow, and that this
flow is initiated at the cooling radius. There does not appear to be a
strong correlation between the presence of filaments and the
structural properties of the BCG, suggesting that the H$\alpha$
filaments are not related to any feedback processes from the central
galaxy. Based on our results, we propose that the gas is cooling out
of the X-ray ICM in clumps at roughly the cooling radius and
collapsing into thin streams as it falls onto the BCG. The observed
asymmetry is likely due to the cooling flow being channeled along
magnetic field lines. Based on the X-ray and H$\alpha$ correlations,
as well as the high effective conductivity in clusters with H$\alpha$
filaments, it is likely that this cool gas is conducting heat from the
surrounding ICM in order to remain ionized. Several questions remain
unanswered in this scenario, such as the cooling process for the hot
gas below $\sim$ 0.5 keV, and the exact ionization mechanism for the
warm gas. Upcoming papers in this series will carefully address these
and other issues.

\section*{Acknowledgements}
Support for this work was provided to M.M., S.V., and D.S.N.R. by NSF
through contract AST 0606932. S.V. also acknowledges support from a
Senior Award from the Alexander von Humboldt Foundation and thanks the
host institution, MPE Garching, where part of this work was
conducted. We thank C.S. Reynolds and S. Teng for useful discussions
and the referee for their careful read of the paper and insightful
suggestions. We thank the technical staff at Las Campanas for their
support during the ground-based observations, particularly David Osip
who helped in the commissioning of MMTF.

\begin{landscape}

\setcounter{table}{0}
\renewcommand{\thetable}{A.\arabic{table}}

\begin{table*}
{\tiny
\caption[]{H$\alpha$, X-Ray, NUV, and radio properties of 23 cooling flow clusters}
\begin{center}
\begin{tabular}{c c c c c c c c c c c c c c c c}
\hline\hline
Name & Fil & F$_{H\alpha,tot}$ & F$_{H\alpha,fil}$ & R$_{H\alpha}$ & Axial & kT$_{100}$ & K$_{30}$ & R$_{cool}$ & \.{M}$_{spec}$ & kT$_{in}$/kT$_{out}$ & Z$_{in}$/Z$_{out}$ & R$_{X,nuc}$ & F$_{NUV}$ &  F$_{NUV}^c$ & F$_{1.4}$ \\
& Type & & & & Ratio & & & & & & & & \\
(1) & (2) & (3) & (4) & (5) & (6) & (7) & (8) & (9) & (10) & (11) & (12) & (13) & (14) & (15) & (16) \\
\hline
\\
Abell~0133    &   I & 0.11(0.03) & 0.03(0.02) &  4.2 & 32.0 &  3.1 &  54 &  50.3 &  0.0 & 2.08/2.54 & 0.95/1.35 & 3.37       & 0.41(0.01) & 0.04(0.09)  &  167\\
Abell~0478    &   I & 2.46(0.25) & 0.83(0.14) & 13.3 & 15.9 &  7.8 &  68 &  58.4 &  0.0 & 5.28/5.75 & 0.66/0.84 & 2.41       & 3.99(2.56) & 3.82(0.04)  &  37\\
Abell~0496    &   I & 2.00(0.13) & 0.81(0.08) & 11.8 & 23.7 &  2.8 &  60 &  55.9 &  1.5 & 1.72/2.52 & 0.60/1.00 & 2.24       & 5.46(0.18) & 4.93(0.12)  &  121\\
Abell~0780    &   I & 1.60(0.15) & 0.16(0.05) & 10.9 & 22.0 &  3.0 &  45 &  61.2 &  7.5 & 3.21/2.99 & 0.66/0.53 & {\bf 0.77} & 3.24(0.03) & 2.84(0.09)  &  40800\\
Abell~1644    &   I & 0.97(0.09) & 0.36(0.05) & 18.1 & 33.3 &  2.9 &  83 &  31.1 &  3.2 & 1.79/2.48 & 0.45/1.34 & 1.82       & 0.97(0.04) & 0.51(0.11)  &  98\\
Abell~1795    &   I & 3.39(0.17) & 1.13(0.10) & 56.0 & 63.7 &  3.9 &  60 &  85.0 &  7.8 & 2.99/4.33 & 0.59/0.53 & 1.43       & 2.64(0.01) & 2.32(0.07)  &  925\\
Abell~2052    &   I & 2.01(0.11) & 0.77(0.07) & 15.5 & 41.4 &  2.2 &  62 &  38.5 &  2.6 & 1.07/1.94 & 0.19/0.77 & {\bf 0.64} & 0.99(0.01) & 0.09(0.21)  &  5500\\
Sersic~159-03 &   I & 1.79(0.15) & 1.06(0.12) & 35.3 & 41.5 &  2.4 &  31 &  58.7 & 13.4 & 1.97/2.42 & 0.42/0.49 & 1.25       & 0.97(0.01) & 0.71(0.06)  &  - \\\\
Abell~0085    &  II & 0.79(0.08) & 0.46(0.06) &  4.1 &  5.9 &  4.0 &  47 &  59.3 &  2.2 & 2.65/2.86 & 0.52/1.48 & 4.11       & 0.56(0.02) & 0.20(0.08)  &  57 \\
Abell~0644    &  II & 0.11(0.03) & 0.02(0.01) &  3.3 &  1.0 &  5.4 & 122 &  71.1 &  1.5 & -/-& -/-&   -        & 0.02(0.05) & -0.27(0.07) &  0\\
Abell~2580    &  II & 0.22(0.04) & 0.01(0.01) &  8.0 &  2.8 &    - &   - &     - &    - & -/-& -/-&   -        & 0.19(0.01) & 0.03(0.04)  &  46\\
Abell~3158    &  II & 0.02(0.01) & 0.00(0.01) &  1.0 &  2.8 &  6.4 & 184 &  43.6 &  0.0 & -/-& -/-& 0.50       & 0.34(0.01) & 0.06(0.07)  &  - \\
Abell~3389    &  II & 0.33(0.06) & 0.02(0.02) &  1.2 &  1.0 &    - &   - &     - &    - & -/-& -/-&   -        & 1.66(0.04) & 0.50(0.27)  &  - \\
Abell~4059    &  II & 0.79(0.08) & 0.00(0.00) &  6.0 &  4.5 &  3.0 &  77 &  51.8 &  0.7 & 1.59/2.04 & 0.49/1.15 & 1.21       & 0.70(0.02) & 0.18(0.12)  &  1280\\
Ophiuchus     &  II & 0.87(0.15) & 0.25(0.08) &  0.1 &  1.0 & 13.4 & 284 &  31.5 &  0.0 & -/-& -/-& 1.63       & 6.56(1.54) & 5.42(0.27)  &  29\\\\
Abell~0970    & III & 0.01(0.01) & 0.01(0.01) &  0.2 &  1.0 &    - &     - &   - &    - & -/-& -/-&   -        &    -(-)    &    -(-)     &  0\\
Abell~1650    & III & 0.03(0.03) & 0.03(0.03) &  0.1 &  1.0 &  5.0 &  91 &  30.5 &  0.0 & -/-& -/-& 1.33       & 0.13(0.01) & -0.12(0.06) &  0\\
Abell~1837    & III & 0.04(0.04) & 0.03(0.03) &  0.3 &  1.0 &    - &     - &   - &    - & -/-& -/-&   -        & 0.34(0.06) & -0.07(0.10) &  5\\
Abell~2029    & III & 0.03(0.03) & 0.02(0.02) &  0.2 &  1.0 &  5.4 &  67 &  87.1 &  3.4 & -/-& -/-& 2.10       & 0.54(0.01) & 0.17(0.09)  &  528\\
Abell~2142    & III & 0.03(0.03) & 0.02(0.02) &  0.1 &  1.0 &  6.0 &  83 & 101.6 &  1.2 & -/-& -/-&   -        & 0.12(0.01) & -0.02(0.03) &  0\\
Abell~2151    & III & 0.02(0.02) & 0.02(0.02) &  0.8 &  1.0 &  2.2 &  55 &  97.3 &  8.4 & -/-& -/-& 1.74       & 0.51(0.02) & 0.07(0.11)  &  2\\
Abell~3376    & III & 0.01(0.01) & 0.00(0.00) &  1.4 &  1.0 &  4.3 & 130 &  85.7 &  2.1 & -/-& -/-& 0.48       & 0.43(0.02) & 0.22(0.05)  &  261\\
WBL~360-03    & III & 0.03(0.03) & 0.02(0.02) &  0.1 &  1.0 &    - &   - &   - & -    & -/-& -/-&   -        &   -(-)     &    -(-)     &  0\\

\hline

\end{tabular}
\end{center}
(1) Cluster name (2) Filament type, following the convention set out in \S3.1 (3) Total H$\alpha$ flux, in units of 10$^{-14}$ erg s$^{-1}$ cm$^{-2}$ (4) H$\alpha$ flux contained in filaments, in units of 10$^{-14}$ erg s$^{-1}$ cm$^{-2}$ (5) Maximum radius of H$\alpha$ emission, in units of kpc (6) Ratio of filament length to width, where the width is typically taken to be the PSF FWHM ($\sim$ 0.6$^{\prime\prime}$) (7) Average X-ray temperature of the cluster in the inner 100 kpc, in units of keV (8) Specific entropy of the cluster at a radius of 30 kpc, in units of keV~cm$^2$ (9) Cooling radius of cluster, in units of kpc (10) Spectrally-determined cooling flow rate, in units of M$_{\odot}$ yr$^{-1}$ (11) Ratio of X-ray temperatures of the gas coincident with the H$\alpha$ filaments to the surrounding ICM derived from a single-temperature plasma model (12) Ratio of X-ray metal abundance of the gas coincident with the H$\alpha$ filaments to the surrounding ICM derived from a single-temperature plasma model (13) FWHM of the X-ray nucleus in units of arcseconds (point source has R$_{X,nuc}$ $\sim$ 0.7$^{\prime\prime}$). Nucleii which are consistent with point sources are shown in bold. (14) Total GALEX NUV flux coincident with the BCG, in units of 10$^{-27}$ erg s$^{-1}$ cm$^{-2}$ Hz$^{-1}$ (15) Total GALEX NUV flux coincident with the BCG, with the contribution from old stars removed (following \S2.2), in units of 10$^{-27}$ erg s$^{-1}$ cm$^{-2}$ Hz$^{-1}$ (16) Radio flux, from the NVSS, in units of mJy.
}

\label{supertable}
\end{table*}
\end{landscape}

\newpage
\appendix 
\subsection*{Properties of Individual Clusters}
In this section, we comment on the clusters in our sample for which
the H$\alpha$ emission is interesting in terms of its flux (or lack
thereof) or morphology. We discuss possible formation mechanisms and
implications on current theories as they apply to each case.
\subsubsection*{Abell~0085}
A large ring can be seen in the X-ray image south of the cluster core,
resembling the outer edge of a bubble. Unlike Abell~2052, however,
this X-ray edge has no H$\alpha$ counterpart. This cluster is one of
three which has asymmetric H$\alpha$ emission on small scales
only. Based on the high H$\alpha$/NUV ratio, it is possible that the
gas could be ionized by a recent burst of star formation in the BCG
center. Previous works (e.g.\ Durret \etal 2005) have suggested that
this cluster has undergone intese merging activity, based on the
temperature/metallicity maps from XMM-Newton. This scenario is
consistent with the relatively high temperature and entropy that we
measure, compared to clusters with more filamentary H$\alpha$
emission.
\subsubsection*{Abell~0133}
A single, thin H$\alpha$ filament extends northeast from the center of
this cluster for $\sim$ 25 kpc.  This filament is only barely
detected, but the detection significance is increased slightly by the
exact coincidence with an X-ray filament. This cluster appears to be
very similar to Abell~1795 in both morphology (long, thin strands in a
single direction) and ionization mechanism (NUV/H$\alpha$ ratio is
consistent with ongoing star formation). Based on XMM-Newton
observations, Fujita \etal (2004) came to the conclusion that this
X-ray filament is likely due to a buoyant radio bubble, since it was
the only scenario put forward that did not directly oppose their
results. However, we show that the metallicity in the filament is
consistent with coming from larger radii, \emph{not} smaller radii,
therefore it is inconsistent with the buoyant radio bubble scenario. 
\subsubsection*{Abell~0478}
The X-ray morphology and H$\alpha$ morphology match up very well in
this cluster, as seen in Figure \ref{bigfig} and also the unsharp masked
X-ray image in Sanderson \etal (2005). There appears to be two nuclei
in both the H$\alpha$ and X-ray emission, while the continuum image is
that of a normal, undisturbed elliptical galaxy. The available UV data
for this cluster is quite poor, but it appears that there is a
significant amount of on-going star formation in the center of this
cluster. However, the combination of shallow data and high extinction
(E(B-V)=0.52) makes it difficult to trust these data. Without a deeper
UV image of the cluster core, it is difficult to determine the
heating source. This cluster also has the highest H$\alpha$ luminosity
of any other in our sample and a significant amount of structure
extending beyond the optical radius of the galaxy, making it one of
the most intriguing clusters in our sample. 
\subsubsection*{Abell~0496}
This cluster has arguably the most interesting H$\alpha$ morphology of
any cluster in our sample. There are at least 5 distinct filaments,
with various shapes and directions. The two longest filaments run
parallel to each other for $\sim$ 12 kpc. Using unsharp mask HST
images, Hatch \etal (2007) found significant structure in the dust
which may be coincident with some of our detected H$\alpha$
filaments. Dupke \etal (2007) also discuss the presence of a cold
front, as can be clearly seen in Figure \ref{bigfig}, which may be due
to an off-center encounter with a massive dark matter halo roughly 0.5
Gyr ago. Regardless of the method of producing this morphology in the
X-ray, we note that the H$\alpha$ filaments do not extend beyond the
cold front in this cluster ($\sim$ 15 kpc), despite the fact that the
cooling radius is nearly 50 kpc.
\subsubsection*{Abell~0780 (Hydra A)}
Wise \etal (2007) present a summary of the X-ray and radio properties
of this cluster, showing the excellent correlation between the radio
jets and the X-ray cavities. The arcing H$\alpha$ filament that we
detect north of the BCG appears to be spatially correlated with the
radio jet. The NUV emission for this cluster is very bright, but it is
unclear if it is has morphological similarities to the H$\alpha$
emission without higher resolution UV imaging. A very bright star to
the west of the BCG made the data reduction slightly more complicated
for this cluster, leading to a lower S/N in the region with the
southern filaments.
\subsubsection*{Abell~1644}
An arcing trail of cool gas extends south from the core of this
cluster. This morphology is likely the result of an off-axis encounter
with a subcluster (Reiprich \etal 2004). We detect a curved H$\alpha$
filament coincident with the cool x-ray gas, as well as two
perpendicular H$\alpha$ strands to the west and north of the cluster
core. These previously-undetected filaments have large length-width
ratios, suggesting that turblence must be low or the magnetic field
strength must be high in this cluster core. Additionally, we find very
little evidence for an AGN in this cluster, in terms of the radio
power, X-ray morphology and hard X-ray flux.
\subsubsection*{Abell~1795}
A pair of thin ($\sim$ 100pc), intertwined filaments extend south from
the cluster core for $\sim$ 50 kpc (MV09). These filaments appear to
be heated by strands of young stars which lie along the full extent of
both filaments. However, the elevated [NII]/H$\alpha$ ratios found in MV09
and the literature suggest that either the star formation is dominated
by massive stars, or another ionizing mechanism is also
important. These twin filaments are coincident with a giant X-ray
filament which is unresolved by Chandra. Additionally, this cluster
has powerful radio jets which are offset in angle from the filament
direction. The cooling wake hypothesis may be valid in this case, as
the BCG appears to have a strong peculiar velocity relative to the
cluster (Oegerle \& Hill 2001). However, we note that a cooling wake
should intuitively have slightly more turbulence than a straight
cooling flow, which is inconsistent with the picture of long, thin
filaments that we observe.
\subsubsection*{Abell~2029}
This cluster is the most massive in our sample and also has the most
massive BCG. However, despite being exceptionally bright at X-ray, NUV
and 1.4 GHz, we detect no H$\alpha$ emission. A clue as to why this
may be lies in the unsharp mask image of the X-ray gas. Unlike most
other massive clusters, Abell 2029 is quite symmetric with no signs of
any structure in the X-ray image. This suggests that the X-ray
structure which we observe in all H$\alpha$-bright clusters is
necessary to produce the filamentary H$\alpha$ emission. This provides
yet another link between the H$\alpha$ filaments and the cooling X-ray
gas and, at the same time, offers a counter-argument to the idea of
buoyant radio bubbles being responsible for the emission.
\subsubsection*{Abell~2052}
The H$\alpha$ emission surrounding the BCG in this cluster is
coincident with radio-blown bubbles in the central region of the
cluster. These bubbles to the north and south of the cluster core are
filled with radio emission, which likely originated from the AGN
within the BCG (Blanton \etal 2003). Since the H$\alpha$ emission is
seen primarily along the edges of the northern bubble, we suspect that
shocks may be responsible for the heating in this case. The very low
NUV/H$\alpha$ ratio seen in Figure \ref{uvha} further supports this
hypothesis. As this cluster is more strongly influenced by the AGN
than any other cluster in our sample, it is an outlier in some of our
plots. As an example, in Figure \ref{abundances} Abell 2052 is represented
by the point with the lowest on-filament metal abundance. This
abundance measurement is most consistent with the core abundance, not
the abundance at the cooling radius. Thus it is possible that our
simple picture of the H$\alpha$ filaments being tied to the cooling
flow may not apply to this cluster.
\subsubsection*{Abell~4059}
The core of this cluster contains two peaks, only one of which is
reproduced in H$\alpha$. The H$\alpha$ emission extends for a short
distance to the west, coincident with an extended arm in the X-ray
image (Choi \etal 2004). The radio emission from this cluster extends
north, perpendicular to the direction of H$\alpha$ emission,
suggesting that the radio jets are not playing a role in the heating.
\subsubsection*{Sersic~159-03}
The optical emission in this cluster has been observed by
Jaffe \etal (2005) and more recently by Oonk \etal (2010). Both
authors find molecular gas which traces the ionized gas out to 20
kpc. We are able to detect H$\alpha$ to 35 kpc, where the surface
brightness blends into the background. It is likely that a deeper
exposure would detect the ionized gas out to the cooling radius of
$\sim$ 60 kpc. There is no evidence for an AGN in the core of this
cluster, despite the fact that it has some of the most extended and
complex filaments. This suggests that the phenomenon of warm ionized
filaments in clusters is likely independent of the presence of an AGN.


\end{document}